\documentclass[twocolumn,aps,prb,longbibliography,superscriptaddress]{revtex4-2}
\usepackage[utf8]{inputenc}
\usepackage[english]{babel}

\usepackage{amsmath}
\usepackage{float}
\usepackage{graphicx}
\usepackage{color}
\usepackage{xcolor}
\usepackage{array}
\usepackage{csquotes}

\begin{document}
\title{Quasimolecular electronic structure of the trimer iridate Ba$_4$NbIr$_3$O$_{12}$}
 
\author{M. Magnaterra}
\affiliation{Institute of Physics II, University of Cologne, 50937 Cologne, Germany}
\author{A. Sandberg}
\affiliation{Department of Physics, Stockholm University, AlbaNova University Center, SE-106 91 Stockholm, Sweden}
\author{H. Schilling}
\affiliation{Sect.\ Crystallography, Institute of Geology and Mineralogy, University of Cologne, 50674 Cologne, Germany}
\author{P. Warzanowski}
\author{L. P\"atzold}
\author{E. Bergamasco}
\affiliation{Institute of Physics II, University of Cologne, 50937 Cologne, Germany}
\author{\mbox{Ch.\ J. Sahle}}
\author{B.~Detlefs} 
\author{K. Ruotsalainen}
\affiliation{European Synchrotron Radiation Facility, BP 220, F-38043 Grenoble Cedex, France}

\author{M. Moretti Sala}
\affiliation{Dipartimento di Fisica, Politecnico di Milano, I-20133 Milano, Italy}
\author{G. Monaco}
\affiliation{Dipartimento di Fisica e Astronomia \enquote{Galileo Galilei}, Universit\`{a} di Padova, Padova, Italy}
\author{P.~Becker}
\affiliation{Sect.\ Crystallography, Institute of Geology and Mineralogy, University of Cologne, 50674 Cologne, Germany}

\author{Q. Faure}
\affiliation{Laboratoire L\'{e}on Brillouin, CEA, CNRS, Universit\'{e} Paris-Saclay, CEA-Saclay, 
	91191 Gif-sur-Yvette, France}
\author{G.S. Thakur}
\affiliation{Department of Chemical Sciences, Indian Institute of Science Education and Research, Berhampur, Odisha, 760003, India}
\affiliation{Max Planck Institute for Chemical Physics of Solids, 01187 Dresden, Germany}
\author{M.~Songvilay}
\affiliation{Universit\'e Grenoble Alpes, CNRS, Institut N\'{e}el, 38042 Grenoble, France}
\author{C. Felser}
\affiliation{Max Planck Institute for Chemical Physics of Solids, 01187 Dresden, Germany}

\author{P.H.M.~van Loosdrecht}
\affiliation{Institute of Physics II, University of Cologne, 50937 Cologne, Germany}
\author{J. van den Brink}
\affiliation{Institute for Theoretical Solid State Physics, IFW Dresden, 01069 Dresden, Germany}
\affiliation{Institute for Theoretical Physics and W\"urzburg-Dresden Cluster of Excellence ct.qmat, Technische Universit\"at Dresden, 01069 Dresden, Germany}
\author{M.~Hermanns}
\affiliation{Department of Physics, Stockholm University, AlbaNova University Center, SE-106 91 Stockholm, Sweden}
\affiliation{Stockholm University, SE-106 91 Stockholm, Sweden}

\author{M. Gr\"uninger}
\affiliation{Institute of Physics II, University of Cologne, 50937 Cologne, Germany}

\begin{abstract}
The insulating mixed-valent Ir$^{+3.66}$ compound Ba$_4$NbIr$_3$O$_{12}$ hosts two holes per Ir$_3$O$_{12}$ trimer unit. We address the electronic structure via resonant inelastic x-ray scattering (RIXS) at the Ir $L_3$ edge and exact diagonalization. The  holes occupy quasimolecular orbitals that are delocalized over a trimer. This gives rise to a rich intra-$t_{2g}$ excitation spectrum that extends from 0.5\,eV to energies larger than 2\,eV.\@ Furthermore, it yields a strong modulation of the RIXS intensity as a function of the transferred momentum {\bf q}. A clear fingerprint of the quasimolecular trimer character is the observation of two modulation periods, $2\pi/d$ and $2\pi/2d$, where $d$ and $2d$ denote the intratrimer Ir-Ir distances. We discuss how the specific modulation reflects the character of the wavefunction of an excited state. 
Our quantitative analysis shows that spin-orbit coupling $\lambda$ of about 0.4\,eV is decisive for the character of the electronic states, despite a large hopping $t_{a_{1g}}$ of about 0.8\,eV.\@ The ground state of a single trimer is described very well by both holes occupying the bonding $j$\,=\,1/2 orbital, forming a vanishing quasimolecular moment with $J$\,=\,0. 
\end{abstract}

	\date{October 31, 2024}   
	\maketitle

\section{Introduction}
In the pursuit of discovering novel magnetic phases, particularly quantum spin liquids, a promising approach is the engineering of unusual interactions between neighboring magnetic moments. An intriguing way to achieve this is through first engineering the character of the local moments themselves. This is possible in cluster Mott insulators -- a novel class of transition-metal compounds where electronic degrees of freedom are delocalized only over individual, small clusters such as dimers, trimers, or tetramers \cite{Khomskii_Rev_2021,Pocha2000,AbdElmeguid2004,Sheckelton2012,Chen2014,Lv2015,Revelli2019resonant,Revelli2022,Magnaterra2023rixs,Magnaterra2024,Jayakumar2023,Chen2024}. The characteristics of the emerging quasimolecular magnetic moments are determined by a complex interplay of spin and orbital degrees of freedom, Coulomb interactions, and intracluster hopping  that depends on the 
cluster shape. 
For instance, the magnetic moments were predicted to be effectively temperature dependent and in particular anisotropic in iridate dimers \cite{Li2020soft}. In general, the unconventional quasimolecular character of the moments is expected to enable magnetic exchange between neighboring clusters that significantly deviates from the conventional Heisenberg exchange, notably permitting bond-direction-dependent exchange interactions.

In a similar fashion, different types of exchange interactions have been realized in, e.g., 
$5d^5$ iridates with $t_{2g}^5$ configuration, 
where spin-orbit entangled $j$\,=\,1/2 moments emerge due to strong spin-orbit coupling. 
In fact, exchange couplings depend strongly on the bonding geometry of, e.g., IrO$_6$ octahedra. 
They vary from isotropic Heisenberg exchange for 180$^\circ$ bonds in corner-sharing configuration to Ising coupling on 90$^\circ$ bonds in edge-sharing geometry \cite{Jackeli09}. The latter allows for the realization of bond-directional Kitaev exchange on tricoordinated lattices \cite{Jackeli09,Chun2015}. Bond-directional magnetic excitations are a hallmark of bond-directional exchange and have been observed in Na$_2$IrO$_3$ in resonant inelastic x-ray scattering (RIXS) \cite{Magnaterra2023BDE}.

Face-sharing IrO$_6$ octahedra, however, feature a very short Ir-Ir distance of roughly 2.5--2.7\,\AA{} \cite{Doi2004,Shimoda2009,Shimoda2010}. This yields much larger hopping $t$, reaching values of the order of 1\,eV \cite{Khomskii2016role,Revelli2019resonant}. 
Accordingly, the existence of exchange-coupled local moments is questionable 
\cite{Revelli2019resonant,Komleva2020three,Li2020soft}. 
Metallic behavior but also a metal-insulator transition have been observed in, e.g., different polytypes of BaIrO$_3$ with face-sharing octahedra \cite{Cao2000BaIrO3,Zhao2009structural,Terasaki2016,Okazaki2018}.
Insulators are found in the Ba$_3$$M$Ir$_2$O$_9$ family 
hosting Ir$_2$O$_9$ dimers. The Ir valence and hence the number of holes per dimer depends 
on the choice of the $M$ ions. 
RIXS data for $M$\,=\,Ce$^{4+}$, Ti$^{4+}$, and In$^{3+}$ \cite{Revelli2019resonant,Revelli2022,Magnaterra2023rixs} demonstrate that the holes occupy 
quasimolecular orbitals and are fully delocalized over a given dimer.

Spin-orbit coupling has been found to be decisive for the character of these quasimolecular 
dimer orbitals \cite{Revelli2019resonant,Revelli2022,Li2020soft}. 
For, e.g., three holes per dimer in face-sharing geometry, the magnetic moment changes from 
$J_{\rm dim}$\,=\,1/2 to 3/2 with increasing intradimer hopping \cite{Li2020soft}, 
and the spin-liquid candidate Ba$_3$InIr$_2$O$_9$ \cite{Dey2017} has been found to be close 
to this transition \cite{Revelli2022}.
Another example for the versatile character of the novel quasimolecular moments in cluster Mott insulators is found in the lacunar spinel GaTa$_4$Se$_8$ \cite{Kim2014tetra,Jeong2017GaTa4,Magnaterra2024}. 
RIXS data show that the Ta$_4$ tetrahedra host quasimolecular $J_{\rm tet}$\,=\,3/2 moments 
for which the actual wavefunction is governed by spin-orbit coupling and the competition 
of different intra-tetrahedral hopping channels \cite{Magnaterra2024}.

\begin{figure}[tb]
\centering
\includegraphics[width=0.95\columnwidth]{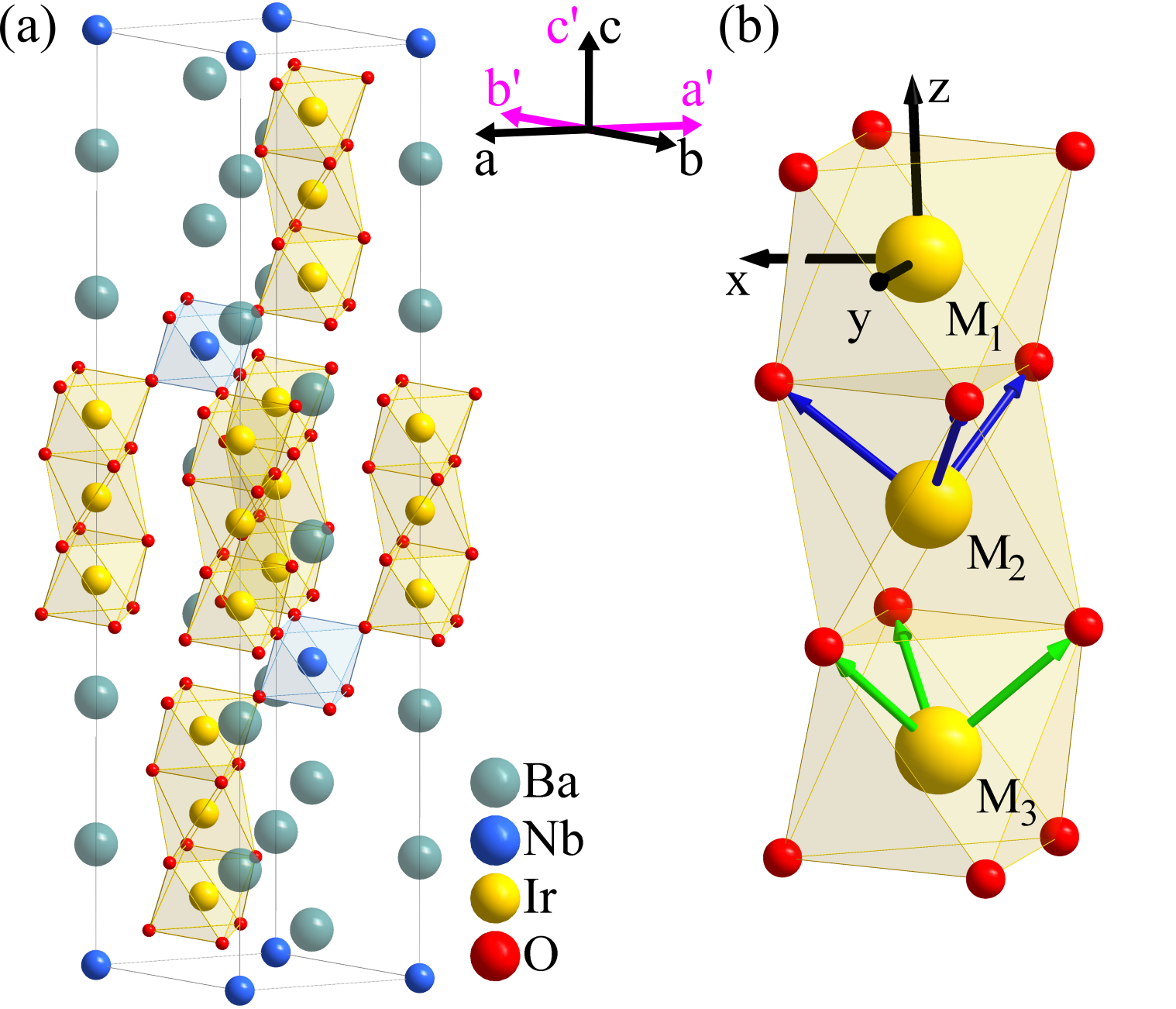}
   \caption{(a) Sketch of the trigonal crystal structure ($R\bar{3}m$) of Ba$_4$NbIr$_3$O$_{12}$ with face-sharing IrO$_6$ octahedra forming well-separated Ir$_3$O$_{12}$ trimer units. The trimer axis is parallel to $c$. 
   The two coordinate frames (black, pink) refer to obverse/reverse twinning.
   (b) An Ir$_3$O$_{12}$ trimer with the three Ir sites $M_i$. The middle site $M_2$ is at the center of inversion and is crystallographically distinct from the outer sites $M_1$ and $M_3$.
    Black: Global coordinate system. 
    Green and blue: Local coordinate systems for the outer and inner octahedra, respectively, with (1,1,1)$_{\rm loc}$ being parallel to the global $c$ axis.
}
\label{fig:struc}
\end{figure}

RIXS is the ideal tool to unravel the quasimolecular electronic structure of 
cluster Mott insulators 
\cite{Ma95,Gelmukhanov94,Revelli2019resonant,Revelli2022,Magnaterra2023rixs,Magnaterra2024,Jeong2017GaTa4,Katukuri2022}. 
The RIXS intensity $I(\mathbf{q})$ of an electronic intra\-cluster excitation at energy $\hbar \omega_0$ reflects the dynamical structure factor $S(\mathbf{q},\omega_0)$, where $\mathbf{q}$ denotes the transferred momentum.
The modulation of $I(\mathbf{q})$ is equivalent to an interference pattern that arises due to 
coherent scattering on all sites of a cluster. Such RIXS interferometry allows us to probe 
the quasimolecular wave function and to uncover the symmetry and character of the electronic 
states, as successfully demonstrated for dimers and tetrahedra
\cite{Revelli2019resonant,Revelli2022,Magnaterra2023rixs,Magnaterra2024}. 
A corresponding modulation of the RIXS intensity can also be studied for, e.g., homonuclear diatomic molecules \cite{Ma95,Gelmukhanov94,Gelmukhanov2021,Soderstrom2024}, bilayer compounds \cite{Porter2022Sr3Ir2}, or the Kitaev magnets Na$_2$IrO$_3$ and $\alpha$-Li$_2$IrO$_3$ with nearest-neighbor spin-spin correlations \cite{Magnaterra2023BDE,Revelli2020finger}.
Furthermore, it has been discussed in the context of witnessing entanglement, using iridate dimers as a model system \cite{Ren2024}. 
For the interpretation of the RIXS data, it is useful to remember that trimers built from face-sharing IrO$_6$ octahedra exhibit inversion symmetry, in contrast to, e.g., face-sharing dimers or tetrahedra. 
The middle site $M_2$ acts as the center of inversion, see Fig.\ \ref{fig:struc}(b). 
All electronic states of a trimer can be classified by their parity. 
Upon inversion, even states are invariant while odd states acquire an overall minus sign. 
The inversion eigenvalue plays a pivotal role for many observables, also for the RIXS intensity $I(\mathbf{q})$. 
We show that $I(\mathbf{q})$ of a trimer behaves qualitatively different for transitions that flip the inversion eigenvalue compared to those that do not. 
As such, RIXS interferometry gives an unambiguous fingerprint of the symmetry of the eigenstates.

Trimer compounds such as Ba$_4$Nb$M_3$O$_{12}$ have been reported for $3d$, $4d$, and $5d$ transition-metal ions 
$M$\,=\,Mn, Ru, Rh, and Ir \cite{Nguyen2019Mn,Nguyen2021,Nguyen2019,Thakur2020,Bandyopadhyay2024gapless,Ali20243dMntrimers,Zhao2024}, 
and a gradual transition from localized electrons for $M$\,=\,Mn to quasimolecular orbitals for $M$\,=\,Ir 
has been discussed based on \textit{ab initio} calculations \cite{Komleva2020three}. 
For instance Ba$_4$NbIr$_3$O$_{12}$ adopts a hexagonal-perovskite structure with well-separated Ir$_3$O$_{12}$ trimers that form triangular layers, see Fig.\ \ref{fig:struc}(a). 
Particularly rich physics ranging from a quantum spin liquid to 
heavy-fermion strange metal behavior has been claimed for 
Ba$_4$Nb$_{1-x}$Ru$_{3+x}$O$_{12}$ \cite{Zhao2024}.
Iridate trimers have been reported with different crystal structures and different hole counts \cite{Shimoda2009,Shimoda2010,Nguyen2019,Nguyen2021,Thakur2020,Miiller2013,Blanchard2016,Blake1999,Ye2018covalency,Volkov2020,Cao2020,Shen2022,Chen2021,Bandyopadhyay2024gapless}. 
Shimoda \textit{et al}.\ synthesized polycrystalline Ba$_4$$Ln$Ir$_3$O$_{12}$ with different $Ln^{3+}$ and $Ln^{4+}$ ions, giving rise to four and three $t_{2g}$ holes per trimer, respectively \cite{Shimoda2009}. 
Based on measurements of the magnetic susceptibility $\chi$, 
the paramagnetic behavior of the $Ln^{3+}$ compounds with four holes per trimer entirely results from the $Ln^{3+}$ ions while the Ir trimers are found to be nonmagnetic \cite{Shimoda2009}. 
In contrast, the three-hole trimers ($Ln$\,=\,Ce$^{4+}$, Pr$^{4+}$, Tb$^{4+}$) order antiferromagnetically at low temperature. 
An interesting variant is realized in Ba$_4$BiIr$_3$O$_{12}$ for which the highly unusual Bi$^{4+}$ valence has been discussed in connection with a  magneto-elastic effect \cite{Miiller2013,Blanchard2016}.  
Three-hole trimers are also realized in Ba$_5$CuIr$_3$O$_{12}$ \cite{Blake1999,Ye2018covalency,Volkov2020} and 
Ba$_4$Ir$_3$O$_{10}$ \cite{Cao2020,Shen2022,Chen2021}. In the latter, adjacent trimers are connected in corner-sharing geometry suggesting large exchange couplings, and a quantum spin liquid state persisting down to 0.2\,K with a very large frustration parameter has been claimed \cite{Cao2020}. RIXS data of Ba$_4$Ir$_3$O$_{10}$ have been interpreted in terms of a spinon continuum extending up to 0.2\,eV \cite{Shen2022}. However, also the occurrence of antiferromagnetic order below $T_N$\,=\,25\,K has been reported \cite{Chen2021}, while 2\,\% Sr substitution yields $T_N$\,=\,130\,K \cite{Cao2020}. 
Remarkably, very different points of view have been reported for these three-hole trimers. The physics of Ba$_4$Ir$_3$O$_{10}$ has been discussed in terms of exchange-coupled $j$\,=\,1/2 moments \cite{Cao2020,Shen2022}, 
while a molecular-orbital picture with a covalency-driven collapse of spin-orbit coupling has been claimed for Ba$_5$CuIr$_3$O$_{12}$ \cite{Ye2018covalency}.

In Ba$_4$NbIr$_3$O$_{12}$ with nonmagnetic $4d^0$ Nb$^{5+}$ ions, the average formal Ir valence amounts to 3.66 for the stoichiometric compound, which corresponds to two $t_{2g}$ holes per Ir trimer. 
Also this compound has been discussed as a spin-liquid candidate \cite{Nguyen2019,Thakur2020,Bandyopadhyay2024gapless}. 
Here, we address the quasimolecular structure via RIXS and exact diagonalization. 
We show how the $\mathbf{q}$-dependent RIXS intensity entails the specific quasimolecular trimer character of a given excited state. Our central result is that the electronic structure of Ba$_4$NbIr$_3$O$_{12}$ qualitatively can be understood in a picture of quasimolecular trimer orbitals built from spin-orbit-entangled $j$ states, in close analogy to the case of face-sharing dimers in Ba$_3$$M$Ir$_2$O$_9$ \cite{Revelli2019resonant,Revelli2022,Magnaterra2023rixs}.
For a single two-hole trimer, we find a nonmagnetic ground state for any realistic set of parameters.

The paper is organized as follows. 
Section \ref{sec:Experimental} addresses experimental aspects such as crystal growth, 
characterization, and RIXS measurements. In Sec.\ \ref{sec:RIXS} we present our RIXS data, 
while Sec.\ \ref{sec:theory} is devoted to the theoretical analysis of RIXS of two-hole trimers. 
We introduce the Hamiltonian and, for an intuitive picture, discuss the single-particle states. Furthermore, we analyze the generic properties of the RIXS response, highlighting the prominent role of inversion symmetry. 
Finally, we compare theory and experiment in Sec.\ \ref{sec:comparison}. 
Beyond the excitation spectrum, we address the $J$\,=\,0 ground state in Sec.\ \ref{sec:groundstate} and the emergence of a small contribution to the magnetic susceptibility in an applied magnetic field in Sec.\ \ref{sec:singlet}.

\section{Experimental}
\label{sec:Experimental}
	
We collected RIXS data on two different single crystals of Ba$_4$Nb$_{1-x}$Ir$_{3+x}$O$_{12}$. Sample A has been grown in Cologne, sample B in Dresden. The RIXS data measured on the two crystals are in excellent agreement with each other (see below). 
The growth of the Dresden crystal B has been described in Ref.\ \cite{Thakur2020}. 
For sample A, we first prepared polycrystalline Ba$_4$NbIr$_{3}$O$_{12}$ using conventional solid-state reaction. A stoichiometric mixture of BaCO$_3$, Nb$_2$O$_5$ and IrO$_2$ was ground and heated in an alumina crucible at 1100$^\circ$C for 48\,h, similar to the method given in Ref.\ \cite{Nguyen2019}. The resulting Ba$_4$NbIr$_{3}$O$_{12}$ was identified as being single-phase by powder x-ray diffraction.
Then, single crystals of Ba$_4$Nb$_{1-x}$Ir$_{3+x}$O$_{12}$ were grown using a flux method inspired by Ref.\ \cite{Thakur2020}. The pre-reacted polycrystalline compound was mixed with BaCl$_2 \cdot 2$H$_2$O in a 1:30 molar ratio, heated in an alumina crucible to 1100$^\circ$C, and slowly cooled down to 950$^\circ$C at a rate of 1$^\circ$C/h. Remains of BaCl$_2$ were dissolved with distilled water.

The isostructural ruthenate Ba$_4$Nb$_{1-x}$Ru$_{3+x}$O$_{12}$ has been reported to change from metallic to insulating behavior as a function of the Nb concentration \cite{Zhao2024}. 
For the iridate Ba$_4$Nb$_{1-x}$Ir$_{3+x}$O$_{12}$ grown in Dresden, the resistivity reveals insulating behavior with an activation energy of 43\,meV \cite{Thakur2020}. We thoroughly addressed the Nb-Ir ratio of our samples. 
For crystals from Dresden, energy-dispersive x-ray spectroscopy (EDX) shows a Nb-Ir ratio 1/3.3, i.e., $x$\,$\approx$\,0.1, while the analysis of single-crystal x-ray diffraction points to $x$\,$\approx$\,0.2 with both Nb$^{5+}$ and excess Ir$^{5+}$ ions occupying the $3a$ site \cite{Thakur2020} that connects adjacent trimers and shows 100\,\% Nb occupation in ideal Ba$_4$NbIr$_{3}$O$_{12}$, see Fig.\ \ref{fig:struc}(a).
The $5d^4$ Ir$^{5+}$ ions are expected to yield nonmagnetic $J$\,=\,0 moments \cite{Yuan2017,Kusch2018,Nag2018,Aczel2022,Warzanowski2023}. 
For crystals grown in Cologne, the chemical composition and homogeneity have been determined using an electronbeam microprobe, see Appendix \ref{App:sample} for details.
The average Nb-Ir ratio is approximately 1/3.4, close to the value found on the Dresden crystals. 
In some crystals, the image of the back-scattered electron detector (BSE) of the microprobe revealed well-separated Ir inclusions and intergrowths with BaIrO$_3$. We selected several crystals with mostly no inclusions or intergrowth and 
measured the Nb-Ir ratio via EDX.\@ Sample A, studied in RIXS, shows a Nb-Ir ratio 
0.97/3.06 that is very close to the nominal value 1/3 of the stoichiometric compound Ba$_4$NbIr$_3$O$_{12}$. 
As discussed below, our RIXS results on samples A and B are fully equivalent 
and do not show any clear contribution of nonmagnetic $5d^4$ Ir$^{5+}$ $J$\,=\,0 ions on Nb sites that may arise in case of off-stoichiometry or Nb-Ir site disorder.

At 300\,K, Ba$_4$Nb$_{1-x}$Ir$_{3+x}$O$_{12}$ exhibits a trigonal crystal structure with space group $R\bar{3}m$ and lattice constants 
$a$\,=\,5.7733(6)\,\AA\ and $c$\,=\,28.720(5)\,\AA\ (sample A). The intratrimer Ir-Ir distance is $d$\,=\,2.54\,\AA. The Ir$_3$O$_{12}$ trimers are oriented parallel to $c$, see Fig.\ \ref{fig:struc}(a). 
Sample A (B) shows an area of about 0.6\,mm $\times$ 0.3\,mm 
(2\,mm $\times$ 1\,mm) perpendicular to the $c$ axis and roughly 0.1\,mm (0.3\,mm) along $c$. 
On both samples, RIXS data were collected on the (001) surface with the (110) direction in the horizontal scattering plane. The quasimolecular trimer character gives rise to a periodic modulation of the RIXS intensity as a function of the transferred momentum $\mathbf{q}$. The period depends on the Ir-Ir distance $d$. Since $d$ is incommensurate with the lattice constant $c$, we use absolute units for $\mathbf{q}$ while still using reciprocal lattice units for ($h$\,\,$k$\,\,$l$).

We performed Ir $L_3$-edge RIXS measurements at beamline ID20 at the ESRF at 300\,K \cite{Moretti2018}. 
The synchrotron was operated with a ring current of about 65\,mA (16-bunch mode), which is roughly 1/3 of the maximum current. The beamline offers two equivalent spectrometers in Johann geometry with 1 or 2\,m Rowland circle diameter \cite{Moretti2018}. 
With the 1\,m spectrometer, the overall intensity of inelastically scattered photons is roughly a factor 4 larger compared to the 2\,m version that offers a better energy resolution. 
To compensate for the reduced ring current that leads to a reduced photon flux, we used the 1\,m Rowland circle spectrometer employing a diced Si(844) Johann crystal (1\,m radius of curvature) with an 80\,mm aperture. 
The resonance behavior has been studied by collecting low-energy-resolution RIXS spectra ($\Delta E$\,=\,0.3\,eV) 
with the incident energy in the range from 11.211 to 11.222\,keV using a Si(311) channel-cut monochromator. 
For all other measurements, we employed a Si(844) backscattering monochromator and set the incident energy to 11.215\,keV to resonantly maximize the RIXS intensity of intra-$t_{2g}$ excitations. For these data, the overall energy resolution was $\Delta E$\,=\,63\,meV as estimated by the full width at half maximum of elastic scattering from a piece of adhesive tape. 
The momentum resolution at, e.g.,  ($h$\,\,$k$\,\,$l$)\,=\,(0.7\,\,0\,\,28.35) 
was $\delta_{\rm hkl}$\,=\,(0.18\,\,0.18\,\,0.8). 
The incident photons were $\pi$ polarized in the horizontal scattering plane.
All RIXS data have been corrected for self-absorption based on the scattering 
geometry \cite{Minola2015}.

\begin{figure}[t]
	\centering
    \includegraphics[width=1\columnwidth]{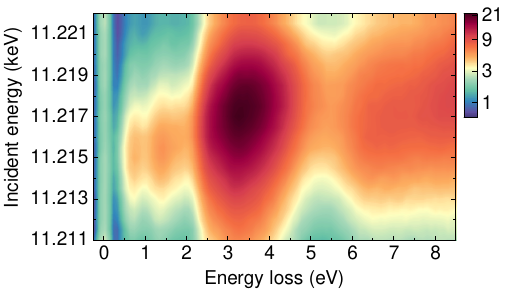}
	\caption{Resonance map of the RIXS intensity of Ba$_4$NbIr$_3$O$_{12}$ (sample A) measured at 300\,K with the low-resolution setup ($\Delta E$\,=\,0.3\,eV) 
    at (0.7\,\,0\,\,36.7). The scattering angle $2\theta$ was close to 90$^\circ$. 
    Excitations to $e_g^\sigma$ state are peaking at an energy loss of about 3.5\,eV for an incident energy $E_{\rm in}$\,=\,11.218\,keV.\@ The intra-$t_{2g}$ excitations at lower energy loss are resonantly enhanced 
    around $E_{\rm in}$\,=\,11.215\,keV.\@ 
    The color scale is logarithmic to highlight the intra-$t_{2g}$ excitations.
}
\label{fig:eg}
\end{figure}

\section{RIXS results}
\label{sec:RIXS}

\begin{figure*}[t]
	\centering
    \includegraphics[width=1\textwidth]{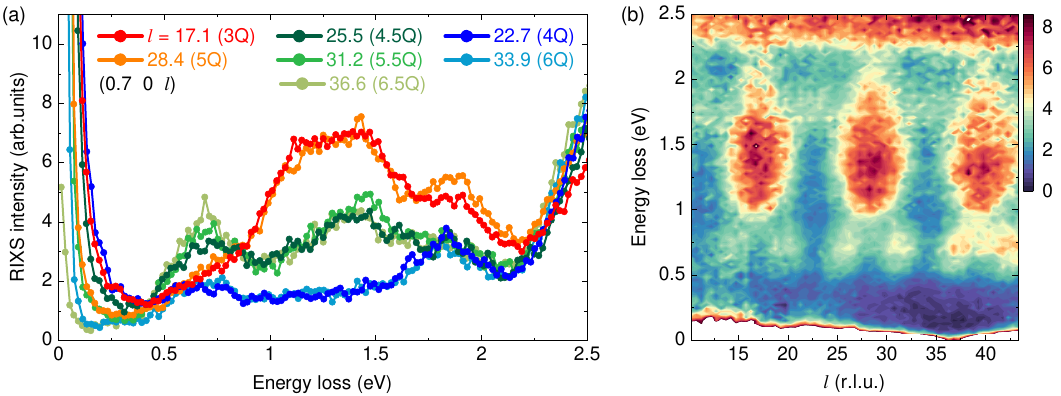}
	\caption{(a) RIXS spectra of Ba$_4$NbIr$_3$O$_{12}$ (sample A) at (0.7\,\,0\,\,$l$) for selected values of $l$ and (b) RIXS intensity map as a function of $l$ at 300\,K.\@ 
    The data do not show any dispersion as a function of the transferred momentum $\mathbf{q}$. Instead, they reveal a pronounced, periodic intensity modulation with two periods, $Q$ and $2Q$, where $Q$\,=\,$\pi/d$\,$\approx$\,$5.67\cdot 2\pi/c$ with the intra-dimer Ir-Ir distance $d$. 
    The data in (b) were merged from measurements at (0.2\,\,-0.2\,\,$l$) for 
    10.25\,$\leq l \leq$\,30.25 and at (0.7\,\,0\,\,$l$) for 20.25\,$\leq l \leq$\,43.25. We used finite $h$ and/or $k$ to avoid enhanced elastic scattering close to a Bragg peak. The very different intensities of the elastic line below about 0.3\,eV reflect the large differences in the scattering angle $2\theta$.
}
\label{fig:Ba3NbIr3O12_spectra_map}
\end{figure*}

The resonance behavior of the RIXS intensity is given in Fig.\ \ref{fig:eg}, 
which shows low-resolution RIXS spectra at ($h$\,\,$k$\,\,$l$)\,=\,(0.7\,\,0\,\,36.7) for incident energies in the range from 11.211 to 11.222\,keV.\@ 
The dominant RIXS feature is observed at an energy loss of about 3.5\,eV.\@ 
It peaks for an incident energy $E_{{\rm in},e_g}$\,$\approx$\,11.218\,keV and corresponds to excitations to $e_g^\sigma$ states. 
The excitation energy of about 3.5\,eV provides an estimate of the cubic 
crystal-field splitting 10\,$Dq$. 
The RIXS peaks at still larger energy loss can be attributed to charge-transfer excitations. 
We focus on the intra-$t_{2g}$ excitations with an energy loss below 2\,eV.\@ 
Their intensity is resonantly enhanced for $E_{{\rm in},t_{2g}}$\,=\,11.215\,keV.\@
The two different resonance energies $E_{{\rm in},e_g}$ and $E_{{\rm in},t_{2g}}$ reflect the x-ray absorption step of the RIXS process, where a $2p$ core electron is promoted to either an $e_g^\sigma$ or a $t_{2g}$ orbital, respectively.

\begin{figure}[t]
		\centering
		\includegraphics[width=0.95\columnwidth]{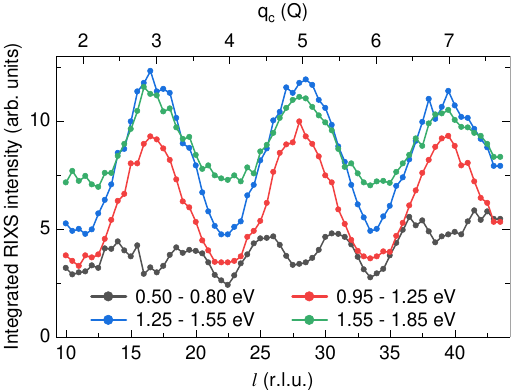}
	\caption{Integrated RIXS intensity as a function of $l$, obtained by integrating the data shown in Fig.\ \ref{fig:Ba3NbIr3O12_spectra_map}(b) for the integration ranges given in the plot.
    With increasing energy, the data have been shifted vertically by 0, 1, 2, and 3, respectively. 
    The top scale uses $Q$\,=\,$\pi/d$ as unit for the component $q_c$ of the 
    transferred momentum. Two periodicities can be identified, $2Q$\,=\,$11.3 \cdot 2\pi/c$ and half of this, $1Q$. 
}
\label{fig:Ba3NbIr3O12_l_integrals}
\end{figure}

Figure \ref{fig:Ba3NbIr3O12_spectra_map}(a) shows RIXS spectra of Ba$_4$NbIr$_3$O$_{12}$ measured with $E_{{\rm in},t_{2g}}$\,=\,11.215\,keV and 63\,meV resolution for transferred momentum (0.7\,\,0\,\,$l$) at selected values of $l$.
The data cover a very large range of $l$ from 17.1 to 36.6, which is possible through the use of hard x-rays at the Ir $L_3$ edge. 
As discussed above (cf.\ Fig.\ \ref{fig:eg}), we attribute the inelastic features between about 0.5 and 2.2\,eV to intra-$t_{2g}$ excitations.
Beyond inelastic features, the spectra in Fig.\ \ref{fig:Ba3NbIr3O12_spectra_map}(a) also show the elastic line at zero loss. Since the data were collected with incident $\pi$ polarization, the elastic line is suppressed for a scattering angle $2\theta$\,=\,90$^\circ$. The data with $l$\,=\,36.6 are closest to this case and accordingly exhibit the smallest elastic line. In contrast, the small value of $2\theta$\,=\,40$^\circ$ for $l$\,=\,17.1  gives rise to a large elastic line that dominates the data up to 0.4\,eV.

Our main experimental result is the pronounced dependence of the RIXS intensity on $l$, i.e., on the component $q_c$ of the transferred momentum parallel to the trimer axis. 
In fact, the RIXS intensity is modulated periodically as a function of $l$, as evident from the color map shown in Fig.\ \ref{fig:Ba3NbIr3O12_spectra_map}(b). 
This provides strong evidence for the quasimolecular character of the excitations \cite{Ma95,Revelli2019resonant,Revelli2022,Magnaterra2024}, as discussed in the introduction and below. 
In particular, we can identify two different periodicities along $l$. In the range from about 1 to 2\,eV, the data show a period $2l_0$\,=\,11.3 (or $2Q$\,=\,$11.3 \cdot 2\pi/c$ in absolute units), while a different intensity modulation with half the period, $1Q$\,$\approx$\,$5.67\cdot 2\pi/c$, is observed around 0.5 -- 0.8\,eV, see Fig.\ \ref{fig:Ba3NbIr3O12_spectra_map}(b). The two different periods can also be seen in Fig.\ \ref{fig:Ba3NbIr3O12_l_integrals}, which shows the RIXS intensity as a function of $l$ for integration over four different energy windows, as given in the plot.
Note that the RIXS spectra in Fig.\ \ref{fig:Ba3NbIr3O12_spectra_map}(a) have been measured at integer and half-integer values of $Q$, highlighting the features for minimum and maximum intensity of the two different modulations. Considering, e.g., the range from 1 to 1.5\,eV and a transferred momentum of $m Q$, maximum intensity is reached for odd $m$ while the intensity minimum is reached for even $m$. 
With the lattice constant $c$\,=\,28.72\,\AA, the long period $2Q$ corresponds to a real-space distance $d$\,=\,$2\pi/2Q$\,=\,2.54\,\AA. This agrees with the nearest-neighbor intratrimer Ir-Ir distance determined by x-ray diffraction, 2.54\,\AA. Accordingly, the short period $1Q$ corresponds to $2d$ in real space, equivalent to the distance between the two outer sites of a trimer. This period $1Q$ is a clear signature of a quasimolecular excitation that mainly involves the two outer trimer sites, as discussed in 
Sec.\ \ref{sec:genericRIXS}.

Qualitatively, the overall features of our RIXS data of Ba$_3$NbIr$_3$O$_{12}$ are very similar to results obtained on the closely related dimer compounds Ba$_3$$M$Ir$_2$O$_9$ with $M$\,=\,Ce and In \cite{Revelli2019resonant,Revelli2022}. First, the spectra exhibit broad peaks with excitation energies that strongly differ from the RIXS response of other iridate Mott insulators with $j$\,=\,1/2 moments localized on individual sites 
\cite{Gretarsson13,Revelli19BaCeIrO6,Magnaterra2023rixs,Warzanowski2024K2IrCl6}. 
Second, the inelastic features show a pronounced modulation of the RIXS intensity that depends on the component of {\bf q} projected onto the trimer axis. 
In contrast to the strong {\bf q} dependence of the intensity, the excitation energy is insensitive to {\bf q}. These properties reflect the localized, quasimolecular nature of the excitations, with charge carriers localized on a given cluster but fully delocalized over the sites of this cluster. 
The actual {\bf q} dependence may serve as a fingerprint of the specific cluster shape. Dimers with Ir-Ir distance $d$ host a sinusoidal interference pattern with period $2Q$\,=\,$2\pi/d$ \cite{Ma95,Revelli2019resonant,Revelli2022,Magnaterra2023rixs}.
The three-dimensional shape of tetrahedral clusters allows for a more complex behavior \cite{Magnaterra2024}. 
The presence of two periods, $1Q$ and $2Q$, is characteristic for a linear trimer with Ir-Ir distances $d$ and $2d$, see Sec.\ \ref{sec:genericRIXS}.
Our RIXS data hence unambiguously demonstrate the quasimolecular character of the electronic structure of Ba$_4$NbIr$_3$O$_{12}$.

\begin{figure}[t]
\centering
 	\includegraphics[width=0.95\columnwidth]{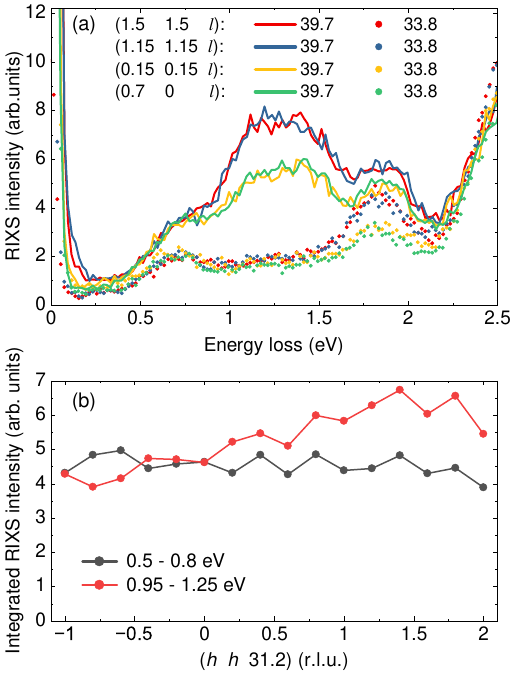}
	\caption{Effect of a change of {\bf q} perpendicular to $c$ (sample A).  
    (a) RIXS spectra at ($h$\,\,$h$\,\,$l$) and (0.7\,\,0\,\,$l$). For each, we compare two values of $l$\,=\,33.8 (dotted) and 39.7 (solid line), equivalent to roughly $6Q$ and $7Q$, respectively. 
    (b) Integrated RIXS intensity along ($h$\,\,$h$\,\,31.2) for two different energy windows. The choice $l$\,=\,31.2, equivalent to $5.5Q$, yields sizable intensity for all RIXS peaks, see Fig.\ \ref{fig:Ba3NbIr3O12_spectra_map}(a). 
    The quasimolecular character of the electronic excitations on a trimer causes 
    a strong modulation as a function of $l$, cf.\ Fig.\ \ref{fig:Ba3NbIr3O12_l_integrals} and Sec.\ \ref{sec:genericRIXS}.
    In contrast, the moderate, non-oscillatory intensity variation as a  
    function of $h$ for $h$\,=\,$k$ with constant $l$ can be attributed to the polarization dependence. In the energy range 0.95 to 1.25\,eV, the maximum intensity in (b) occurs for large $h$ around 1.5, in agreement with the data in (a).
}
\label{fig:Ba3NbIr3O12_hk_integrals}
\end{figure}

For a trimer running along the $c$ axis, the RIXS intensity strongly depends on $l$. In contrast, $h$ and $k$ are expected to play a minor role. Also this is supported by experiment, see Fig.\ \ref{fig:Ba3NbIr3O12_hk_integrals}. Panel (a) shows RIXS spectra for different ($h$\,\,$k$\,\,$l$). For each set of $h$ and $k$, data are given for $l$\,=\,33.8 and 39.7. For $l_0$\,=\,5.67, these $l$ values roughly correspond to $6l_0$ and $7l_0$ (or $6Q$ and $7Q$), respectively. For comparison, panel (b) depicts the integrated RIXS intensity in the energy ranges 0.5 -- 0.8\,eV and 0.95 -- 1.25\,eV for $l$\,=\,31.2 or $5.5Q$. The insensitivity to $h$ and $k$ is particularly evident for the data in Fig.\ \ref{fig:Ba3NbIr3O12_hk_integrals}(a) with $6Q$, i.e., minimum intensity,  
and for the window 0.5 -- 0.8\,eV.\@ Around 1 -- 1.5\,eV, the intensity moderately increases with $h$ and $k$. We emphasize that this slow increase as a function of $h$ and $k$ strongly differs from the oscillating behavior as a function of $l$. This slow increase can be attributed to a polarization effect since the large change of {\bf q} requires a corresponding change of the scattering geometry. Very similar polarization effects as a function of $h$ and $k$ have been observed in iridate dimer compounds \cite{Revelli2022}.

\begin{figure*}[bt]
\centering
  		\includegraphics[width=1\textwidth]{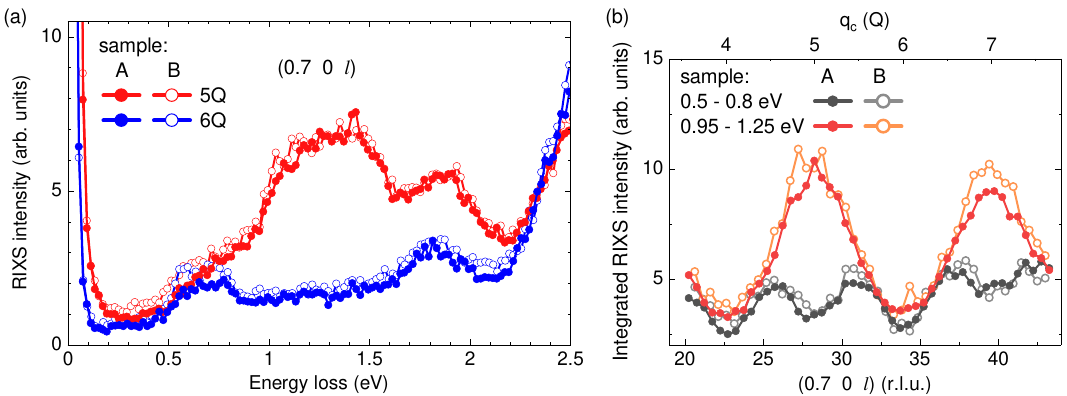}
		\caption{Comparison of (a) RIXS spectra and (b) the $\mathbf{q}$-dependent RIXS intensity of the two different samples A (full) and B (open symbols), see Sec.\ \ref{sec:Experimental}.
		The data agree very well with each other, pointing towards a negligible role of Nb-Ir disorder.
}
\label{fig:Ba3NbIr3O12_samples_comparison}
\end{figure*}

Finally, we address the possible role of Nb-Ir site disorder. We first consider a single $5d^4$ Ir$^{5+}$ ion on a Nb$^{5+}$ site (Wyckoff position $3a$), forming a local $J$\,=\,0 moment. These could be identified by observing two characteristic narrow peaks in the energy range up to 3/2\,$\lambda$ \cite{Yuan2017,Kusch2018,Nag2018,Aczel2022,Warzanowski2023}, 
roughly 0.6 to 0.7\,eV.\@ Similarly, a single $5d^5$ Ir$^{4+}$ ion would cause a sharp feature at 3/2\,$\lambda$. 
The absence of such narrow peaks in our data suggests a minor role of Nb-Ir site disorder. This is supported by the {\bf q} dependence of the RIXS intensity that reflects which Ir sites are involved in a given excitation. It thus can be used to detect signatures of disorder, e.g., excitations related to Ir ions located on the Nb $3a$ site that interact with Ir ions on trimer sites. 
This has been shown for the dimer compound 
Ba$_3$Ti$_{3-x}$Ir$_x$O$_9$ \cite{Magnaterra2023rixs} with a sizable contribution of Ti-Ir site disorder. The Ti sites connect neighboring Ir$_2$O$_9$ dimers in the same way as the Nb $3a$ sites connect the trimers in Ba$_4$NbIr$_3$O$_{12}$, see Fig.\ \ref{fig:struc}(a).  
Considering the projection onto the $c$ axis, the Nb and Ir sites are displaced by 2.24\,\AA, which is smaller than the intratrimer Ir-Ir distance $d$\,=\,2.54\,\AA. Accordingly, Ir$_{3a}$-Ir pairs with one Ir ion on the Nb $3a$ site are expected to cause a modulation with a larger period $l_{\rm Nb}$\,=\,12.8. 
This is not present in our data. In addition, such pairs are also displaced perpendicular to $c$ such that 
corresponding excitations would show a sinusoidal modulation as a function of $h$ or $k$, as discussed below in Sec.\ \ref{sec:genericRIXS}. 
Such a modulation indeed has been observed in Ba$_3$Ti$_{3-x}$Ir$_x$O$_9$ \cite{Magnaterra2023rixs}. 
Figure \ref{fig:Ba3NbIr3O12_hk_integrals}(b) depicts the integrated RIXS intensity along ($h$\,\,$h$\,\,31.2) for two different energy ranges. 
Along this direction, Ir$_{3a}$-Ir pairs would yield an intensity modulation following either
$\sin^2(\pi(h\pm 31.2/12.8))$ or $\cos^2(\pi(h\pm 31.2/12.8))$ with a period of 1 in $h$, in close analogy to the case of Ba$_3$Ti$_{3-x}$Ir$_x$O$_9$ \cite{Magnaterra2023rixs}. However, such modulation as a function of $h$ is not observed in Fig.\ \ref{fig:Ba3NbIr3O12_hk_integrals}(b). We conclude that Nb-Ir site disorder is negligible for the discussion of our RIXS data. 
This is further supported by the comparison of data of the two different samples A and B, grown in Cologne and Dresden, see Sec.\ \ref{sec:Experimental}. Figure \ref{fig:Ba3NbIr3O12_samples_comparison} compares  RIXS spectra of the two samples in panel (a) and the $\mathbf{q}$-dependent RIXS intensity  in panel (b). 
The data of the two samples agree very well with each other.

\section{Theoretical analysis}
\label{sec:theory}

\subsection{Hamiltonian for a single trimer}
\label{sec:Hamiltonian}

In the following, we address the RIXS response of Ba$_4$NbIr$_3$O$_{12}$ below 2\,eV.\@ 
Due to the large cubic crystal-field splitting $10$\,$Dq$, see Sec.\ \ref{sec:RIXS}, 
we restrict the discussion to t$_{2g}$ orbitals. For a trimer with two holes, 
the full Hamiltonian consists of spin-orbit coupling, non-cubic crystal-field splitting,  Coulomb repulsion and Hund's coupling \cite{Perkins_interplay_2014}, as well as hopping between neighboring sites,  
\begin{align}\label{eq:hamiltonian}
	H & = \sum_i \left( H_{{\rm soc},i} + H_{\Delta,i} + H_{{\rm C},i}\right) + H_{t} \, ,
\end{align}
where $i$ runs over the three trimer sites.
The first two terms, $H_{{\rm soc},i}$ and $H_{\Delta,i}$, 
describe the single-ion physics in the case of a single $t_{2g}$ hole per site \cite{Moretti2014CEF}. The spin-orbit coupling term has the form  
  \begin{align} \label{eq:spin-orbit}
	H_{{\rm soc},i} & = \lambda \, \mathbf{L}_i \cdot \mathbf{S}_i,
  \end{align}
where $\lambda$ is the effective coupling constant, 
\textbf{L}$_i$ the effective orbital momentum on site $i$, and \textbf{S}$_i$ the spin. 
Iridium is a heavy atom with large $\lambda$\,=\,0.4--0.5\,eV \cite{Gretarsson13,Revelli2019resonant,Revelli19BaCeIrO6}. 
For an Ir site with a single $t_{2g}$ hole, a positive $\lambda$ splits the $t_{2g}$ orbitals into a low-lying $j$\,=\,1/2 doublet and a $j$\,=\,3/2 quadruplet that is 3/2\,$\lambda$ higher in energy. We use the global $z$ axis as spin quantization axis \cite{Li2020soft},
since the local coordinate systems for the IrO$_6$ octahedra have a different orientation for the middle and outer Ir sites, see Fig.\ \ref{fig:struc}(b).

The second term of the Hamiltonian, the trigonal crystal-field splitting, is given by 
\begin{align} \label{eq:trigonal}
    H_{\Delta,i} & = \Delta_i \cdot L_{z,i}^2 -\overline{E}_{\Delta,i},
\end{align}
where $L_{z,i}$ is the orbital momentum along the global $z$ axis and 
$\overline{E}_{\Delta,i}$ denotes the average energy. 
For the trimer, a trigonal crystal-field splitting $\Delta_i$ 
arises for a corresponding distortion of the IrO$_6$ octahedra. However, 
finite $\Delta_i$ is also caused by the linear arrangement of the Ir neighbors even in the case of perfectly cubic octahedra \cite{Kugel_spin-orbital_2015}.
The $t_{2g}$ orbitals are split into $a_{1g}$ and $e_{g}^\pi$ orbitals with an energy gap $\Delta_i$.
Our convention in the hole picture is such that the $a_{1g}$ orbitals are lower in energy for positive $\Delta_i$.

As the middle ($m$) and outer sites ($o$) are crystallographically distinct, see Fig.\ \ref{fig:struc}(b), we allow for different crystal-field splittings $\Delta_{m}$ and $\Delta_{o}$.
In fact, \textit{ab initio} simulations suggest that the two may differ substantially in magnitude and even have opposite signs \cite{Komleva2020three}. In Eq.\ \eqref{eq:trigonal}, we subtract the average energy 
$\overline{E}_{\Delta_{m}}$ or $\overline{E}_{\Delta_{o}}$ for each site to balance the energies around zero. 
This term is important for $\Delta_m\neq\Delta_o$. Without it, the relative energies of middle and outer sites become incorrect.
For $\Delta_m$\,=\,$\Delta_o$, it only adds an overall energy constant.

On-site Coulomb interactions are described by the third term in the Hamiltonian, Eq.\ \eqref{eq:hamiltonian}, as follows \cite{Perkins_interplay_2014}
\begin{align}
  \begin{aligned}\nonumber
   H_{{\rm C},i} & =  U \, \sum_{\alpha} n_{i\alpha \uparrow} n_{i\alpha \downarrow}
		+  \frac{1}{2} (U-3J_H) \sum_{\sigma,\alpha\neq\alpha^\prime} n_{i\alpha \sigma} n_{i\alpha^\prime \sigma} \\
		& + (U-2J_H) \sum_{\alpha\neq\alpha^\prime} n_{i\alpha \uparrow} n_{i\alpha^\prime \downarrow}
		\\
		& +  (U-2J_H)\left( 15-5\sum_{\alpha,\sigma} n_{i\alpha \sigma} \right) \\
  & +  J_H \sum_{\alpha \neq \alpha'} \left( c^\dagger_{i\alpha \uparrow} c^\dagger_{i\alpha \downarrow} c_{i\alpha'\downarrow} c_{i\alpha'\uparrow}
		- c^\dagger_{i\alpha\uparrow} c_{i\alpha\downarrow} c^\dagger_{i\alpha'\downarrow} c_{i\alpha'\uparrow}\right), 
  \end{aligned}
\end{align}
where $c^\dagger_{i\alpha\sigma}$ ($c_{i\alpha\sigma}$) creates (annihilates) a hole at site $M_i$ with $i$\,=\,1,2,3 of orbital type $\alpha$ with spin $\sigma=\uparrow,\downarrow$, and $n_{i\alpha\sigma}=c^\dagger_{i\alpha\sigma}c_{i\alpha\sigma}$ counts the number of holes. The intraorbital Coulomb repulsion is given by $U$, and $J_H$ denotes Hund's coupling. The third line applies in the hole picture and takes care of the relative energies between different sites.

Finally, the hopping of holes between neighboring Ir sites is described by the last term of the Hamiltonian, 
    \begin{align}
  H_t &= \sum_{\langle i,j\rangle, \alpha, \alpha',\sigma} t_{ij}^{\alpha\alpha'} c^\dagger_{i\alpha\sigma} c_{j\alpha'\sigma}.
	\end{align}
In face-sharing geometry, as a result of the threefold symmetry around the trigonal axis, direct Ir-Ir hopping and ligand-mediated hopping both are diagonal in the $a_{1g}$/$e_{g}^\pi$ basis \cite{Kugel_spin-orbital_2015}. 
In other words, the only non-zero hopping elements are those between the same $a_{1g}$ or $e_g^\pi$ orbitals on neighboring sites. 
Their respective hopping strength is denoted by $t_{a_{1g}}$ and $t_{e_g^\pi}$, respectively, with realistic values for their ratio $f$\,=\,$t_{e_g^\pi}/t_{a_{1g}}$ being close to $f$\,=\,-1/2 \cite{Li2020soft}. We will use $t$\,=\,$t_{a_{1g}}$ and $f$ to parameterize the hopping.

\subsection{Single-particle states}

\begin{figure}[t]
    \centering
    \includegraphics[width=\columnwidth]{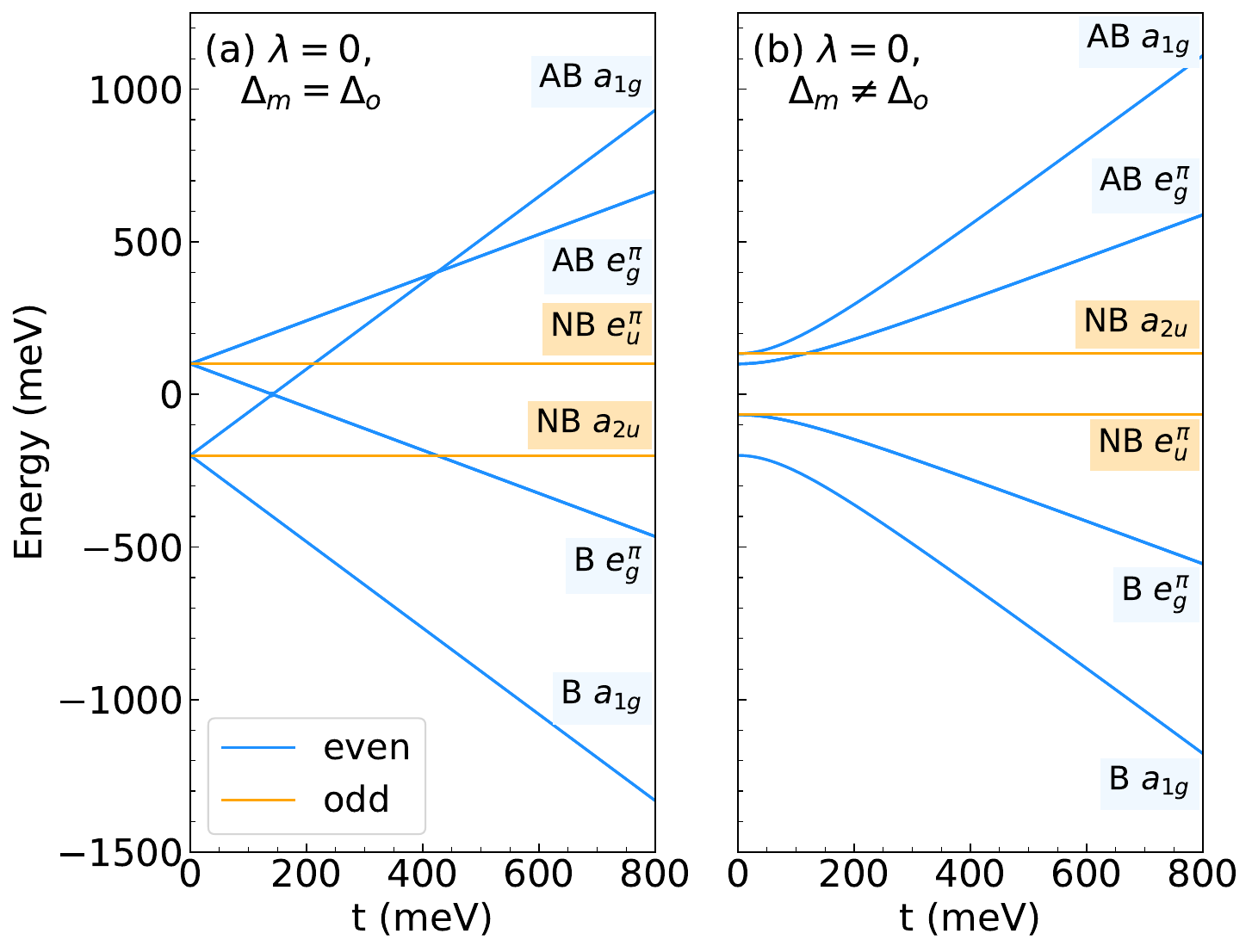}
\caption{Formation of quasimolecular B/NB/AB orbitals from single-site $a_{1g}$ and $e_{g}^\pi$ states as a function of $t$\,=\,$t_{a_{1g}}$ for 
$f$\,=\,$t_{e_g^\pi}/t_{a_{1g}}$\,=\,-1/2 in a single-particle picture with $\lambda$\,=\,0. 
(a) $\Delta_m$\,=\,$\Delta_o$\,=\,0.3\,eV.\@ 
(b) $\Delta_m$\,=\,0.3\,eV, $\Delta_o$\,=\,-0.2\,eV.\@ 
Blue (yellow) denotes even (odd) states. 
}  
\label{fig:single-hole_hopping}
\end{figure}

For an intuitive picture, we consider the non-interact\-ing case, $U$\,=\,$J_H$\,=\,0, and discuss the formation of quasimolecular orbitals in two different limits, 
$\Delta \gg \lambda$ vs.\ $\Delta \ll \lambda$.
First, we address the effect of hopping on the $a_{1g}$ and $e_g^\pi$ orbitals for $\lambda$\,=\,0. 
Due to the very short Ir-Ir distance, we expect hopping to be large. Values of $t_{a_{1g}}$\,=\,0.5--1.1\,eV have been derived from the analysis of RIXS data of the face-sharing dimer compounds Ba$_3$CeIr$_2$O$_9$ and Ba$_3$InIr$_2$O$_9$ \cite{Revelli2019resonant,Revelli2022}.
In the trimer, hopping yields bonding (B), non-bonding (NB), and anti-bonding (AB) orbitals. 
The energy of non-bonding states is not affected by $t$, while the shift of B/AB states in the simplest case, $\Delta_m$\,=\,$\Delta_o$, is linear in $t$, see Fig.\ \ref{fig:single-hole_hopping}(a).
For $\Delta_m \neq \Delta_o$, the hopping Hamiltonian $H_t$ does not commute with the crystal-field term $H_\Delta$, which scrambles the order of states for small $t$ and leads to a more complicated behavior, see Fig.\ \ref{fig:single-hole_hopping}(b). 
For realistic, large values of $t$, however, the qualitative behavior is the same in Figs.\  \ref{fig:single-hole_hopping}(a) and (b).

\begin{figure}[tb]
	\centering
	\includegraphics[width=\columnwidth]{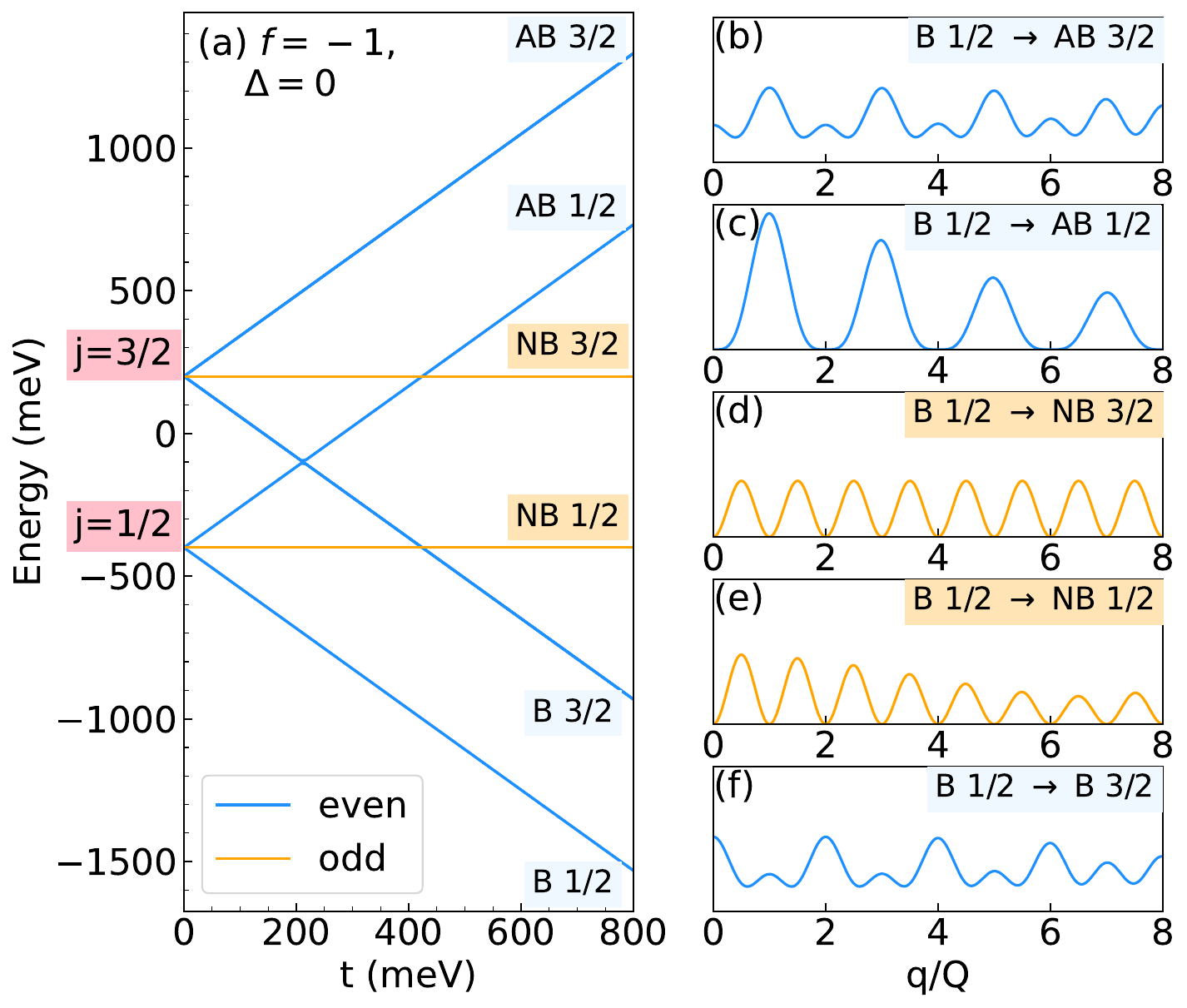}
	\caption{Quasimolecular orbital picture for strong spin-orbit coupling and vanishing trigonal field, $\Delta$\,=\,0. 
    (a) Effect of hopping $t$ on the spin-orbit entangled $j$ states in a single-hole picture for $\lambda$\,=\,0.4\,eV.\@
    The plot depicts the special case with $f$\,=\,-1, for which $j$ remains a good quantum number of the B/NB/AB states. 
    (b)-(f) RIXS intensity for a single hole per trimer 
    with $q$ along the trimer axis and $Q$\,=\,$\pi/d$.
    The five panels show the rich behavior of the $q$ dependence for excitations from the bonding 
    $j$\,=\,1/2 ground state to each of the excited states.  
}
\label{fig:single_hole_I_vs_dq_spin_orbit_summary}
\end{figure}

The opposite limit is achieved for $\Delta_m$\,=\,$\Delta_o$\,=\,0, describing the effect of hopping on the spin-orbit entangled $j$ states. In general, hopping mixes the $j$\,=\,1/2 and 
$j$\,=\,3/2 multiplets \cite{Li2020soft,Revelli2022}.
There is however a special value $f$\,=\,-1, for which $H_t$ and $H_{\rm soc}$ commute such that $j$ remains a good quantum number of the B/NB/AB states, 
see Fig.\ \ref{fig:single_hole_I_vs_dq_spin_orbit_summary}(a). 
Although $f$\,=\,-1 is not realized in Ba$_4$NbIr$_3$O$_{12}$, 
this case is helpful to simplify some of the discussion below.
For $\Delta_m$\,=\,$\Delta_o$\,=\,0 and $f$\,=\,-1, the quasimolecular states are B/NB/AB doublets built from $j$\,=\,1/2 states or B/NB/AB quartets made from $j$\,=\,3/2 states. 
Note that a further limit, considering exchange-coupled local $j$ moments, requires to consider finite $U$. 
However, two holes on three sites remain delocalized even for $t \ll U$ such that this limit cannot be realized here.

In the presence of inversion symmetry, it is easy to verify that the single-particle bonding and anti-bonding states of a trimer are even under inversion, while the non-bonding states are odd. A generic inversion-odd single-particle state can be written as 
\begin{align}
    \label{eq:generic_sp_odd}
    |\psi_{o}\rangle &= \frac{1}{\sqrt{2}}\left(|m\rangle_1 - |m\rangle_3 \right),
\end{align}
where $|m\rangle_i$ is a generic single-particle state on site $M_i$.  
The hopping amplitudes towards the middle site $M_2$ cancel, and hence the energy of an odd state does not depend on (nearest-neighbor) hopping, i.e., odd states are non-bonding.
The odd non-bonding states built from single-site $a_{1g}$ and $e_g^\pi$ orbitals have 
$a_{2u}$ and $e_u^\pi$ character, respectively. 
In contrast a generic inversion-even state can be written as 
\begin{align}
    \label{eq:generic_sp_even}
    |\psi_{e}\rangle &= v |m\rangle_1 +u|n\rangle_2 + v |m\rangle_3,
\end{align}
where $|m\rangle$ and $|n\rangle$ are generic single-particle states and  $u$ and $v$ are complex parameters with $|u|^2+2|v|^2=1$. 
This form of the even/odd states will prove useful in the next section to gain a better understanding of the RIXS intensities for the trimer system.

\subsection{Generic properties of RIXS intensities for inversion-symmetric trimers}
\label{sec:genericRIXS}

To put the RIXS response of trimers into perspective, we briefly recap the behavior of dimers. 
Coherent summation over resonant scattering processes on the two sites of a dimer yields a {\bf q}-dependent modulation of the RIXS intensity. This interference pattern can be described in terms of an inelastic version of Young's double slit experiment \cite{Revelli2019resonant,Ma95}. 
In other words, RIXS probes the dynamical structure factor 
$S(\mathbf{q},\omega_0)$ of a dimer excitation at energy $\hbar \omega_0$ that  exhibits a sinusoidal {\bf q} dependence. 
In the presence of inversion symmetry, the RIXS intensity for a dimer is given by 
\begin{align}
\label{eq:dim}
    I_{eo}^{\rm dim}(q) &\sim \sin^2 (qd/2), 
    \hspace*{8mm}  
    I_{ee}^{\rm dim}(q) \sim \cos^2 (qd/2), 
\end{align}
where $d$ is the distance between the two sites of a dimer, 
$q$ denotes the component of {\bf q} parallel to the dimer axis, 
and $I_{eo}^{\rm dim}$ refers to excitations on the dimer that flip the symmetry from even to odd or vice versa while $I_{ee}^{\rm dim}$ corresponds to excitations from even to even or from odd to odd states. The period of the interference pattern $2Q$\,=\,$2\pi/d$ measures the intradimer distance, while a $\cos^2(qd/2)$ or $\sin^2(qd/2)$ behavior reveals the symmetry and character of the states involved in the excitation. 
A trimer offers similar but richer behavior.

Before we consider RIXS on Ba$_4$NbIr$_3$O$_{12}$, let us first discuss some general properties of inversion-symmetric trimers. The RIXS amplitude $A(\mathbf{q})$ is given by \cite{Ament11,Ament11rev}
	\begin{align}
    \label{eq:RIXS_amplitude}
		A(\mathbf{q})\sim \langle \psi_{f} | \sum_{\mathbf{R}} e^{i \mathbf{q}\mathbf{R}} \left[D^\dagger(\epsilon_{\rm out}^*)D(\epsilon_{\rm in})\right]_\mathbf{R} |\psi_{i}\rangle
	\end{align}
where $\mathbf{R}=(0,0,0), \, (0,0,\pm d)$ for the three sites of the trimer, $D$ is the local dipole transition operator, $|\psi_{i}\rangle$ and $|\psi_{f} \rangle$ 
denote the initial and final state, respectively, 
and $\epsilon_{\rm in}$ and $\epsilon_{\rm out}$ denote the incident and outgoing polarization, respectively. 
The corresponding RIXS intensity is obtained as $I(\mathbf{q})\sim |A(\mathbf{q})|^2$. 
In the experiment, 
$\epsilon_{\rm in}$ was oriented within the horizontal scattering plane. For the total intensity, we sum the intensities for vertical and horizontal outgoing polarization $\epsilon_{\rm out}$. 
For Ba$_4$NbIr$_3$O$_{12}$, we additionally take into account that obverse/reverse twinning is common in rhombohedral structures. The two twins cannot be distinguished in, e.g., a Laue x-ray diffraction image. For the two-hole RIXS intensities discussed in Sec.\ \ref{sec:comparison}, we hence sum over the two possible trimer orientations, rotated by $\pi$ around the $c$ axis, that correspond to the two twin domains. This has only a minor effect on the result. It only affects the polarization and thus the slowly varying envelope of the interference pattern. 
Note that the single-hole RIXS intensities as depicted in Fig. \ref{fig:single_hole_I_vs_dq_spin_orbit_summary} only consider one orientation.

\begin{figure}[tb]
    \centering
    \includegraphics[width=\columnwidth]{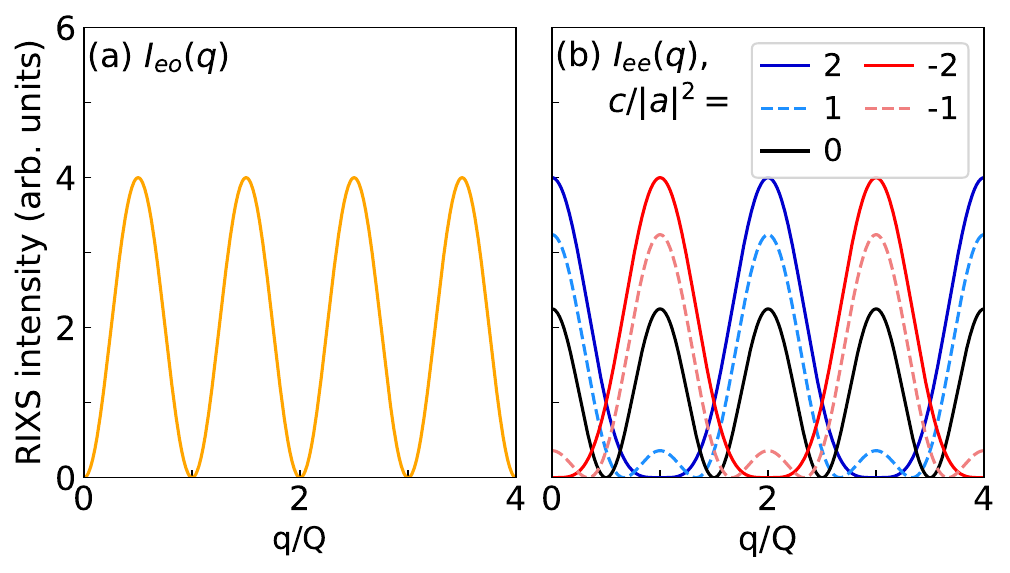}
	\caption{Generic interference patterns for a trimer with inversion symmetry, carrying information on the character and symmetry of the states involved. Note that $q$ is the component of {\bf q} parallel to the trimer axis, with $Q$\,=\,$\pi/d$. 
    (a) $|a_{eo}|^2 \sin^2(qd)$\,=\,$|a_{eo}|^2 \sin^2(\pi q/Q)$ behavior is characteristic for excitations from even to odd states or vice versa, see $I_{eo}(q)$ in Eq.\ \eqref{eq:odd}. 
    (b) The interference between inner and outer sites of the trimer yields a richer behavior for excitations from even to even or odd to odd states, see $I_{ee}(q)$ in Eq.\ \eqref{eq:even}. 
    The plot shows $I_{ee}(q)$ for different $c/|a|^2$, illustrating the 
    qualitatively different behavior of $|c|$\,=\,$2|a|^2$ (blue/red), 
    $|c|$\,=\,$|a|^2$ (dashed blue/dashed red), and $c$\,=\,0 (black). 
    To improve  visibility, we have chosen $a$\,=\,1, 1.2, and 1.5 for 
    $|c/a^2|$\,=\,2, 1, and 0, respectively.
    Here, we neglect the change of polarization with $q$ that adds a slow envelope.
}
\label{fig:I_ee-function}
\end{figure}

For an excitation that flips the symmetry from even to odd or from odd to even, the intensity is particularly simple. 
We will argue for its form using the single-particle expressions in Eqs.\ \eqref{eq:generic_sp_odd} and \eqref{eq:generic_sp_even}, 
but the resulting Eq.\ \eqref{eq:odd} is valid also for the many-body case, see
Appendix \ref{sec:Intensity_inversion}. 
Since an odd state has no occupation on the middle site $M_2$, the amplitude contains only two terms, 
\begin{align} \nonumber
   A_{eo}(q) &\sim \frac{v}{\sqrt{2}} \left[ e^{i q d} \langle \tilde m|D^\dagger D |m\rangle_1 - 
 e^{-i q d} \langle \tilde m|D^\dagger D |m\rangle_3 \right], 
\end{align}
where q denotes the $z$ component, parallel to the trimer axis.
As the sites $M_1$ and $M_3$ are equivalent, their matrix elements are identical, 
which yields
\begin{align}\label{eq:odd}
	I_{eo}(q) &\sim |a_{eo}|^2\sin^2(q d), 
\end{align}
where the parameter $a_{eo}$ depends on 
the initial and final states but also on the scattering geometry and the corresponding polarization, which causes an effective $\mathbf{q}$ dependence of $a_{eo}$. 
For hard x-rays at the Ir $L_3$ edge, covering a large range of {\bf q}, $a_{eo}(\mathbf{q})$ serves as a slow envelope to the sinusoidal modulation. 
Neglecting this envelope, the intensity $I_{eo}(q)$ exhibits a period $Q$\,=\,$2\pi/2d$ since it is blind to the middle site and consequently shows the sinusoidal intensity modulation of an effective `dimer' with distance $2d$, see Eq.\ \eqref{eq:dim} and Fig.\ \ref{fig:I_ee-function}(a). 
Intensity maxima occur for $q$\,=\,$(m+1/2)Q$ with integer $m$. 
This agrees with the experimental result at low energies, see black curve in Fig.\ \ref{fig:Ba3NbIr3O12_l_integrals} and Sec.\ \ref{sec:comparison}.

For an even-to-even excitation, the RIXS intensity of a trimer is more abundant. 
Using the generic form of even states given in Eq.\ \eqref{eq:generic_sp_even} and the equivalence of the matrix elements on the outer sites $M_1$ and $M_3$, the RIXS amplitude can be written as 
\begin{align}
    \label{eq:even-even-Amp}
    A_{ee}(q)&\sim (e^{iqd}\!+\!e^{-iqd})\tilde v^* v \langle \tilde m|D^\dagger  D |m\rangle_3 
    + \tilde u^* u \langle \tilde n|D^\dagger  D |n\rangle_2\nonumber\\
    &\equiv a \cos(qd) + b, 
\end{align}
where $a \cos(qd) $ is the sum of the amplitudes on the outer sites while $b$ is the amplitude on $M_2$. This implies
\begin{align}\label{eq:even}
 I_{ee}(q)&\sim  |a|^2  \cos^2(qd) + c \cos(qd) + |b|^2 
\end{align}
where $c$\,=\,$(a^*\!b\!+\!b^*\!a)$ captures the interference  between the outer sites and the middle site $M_2$.
In this general form, $I_{ee}(q)$ applies to the many-body case for both even-to-even and odd-to-odd excitations, 
see Appendix \ref{sec:Intensity_inversion}.

Note that $a \sim v \tilde v^*$ and $b \sim u \tilde u^*$ reflect the coefficients of the wave functions of initial and final states, see Eqs.\ \eqref{eq:generic_sp_even} and \eqref{eq:even-even-Amp}. 
The rich structure of $I_{ee}(q)$ hence contains substantial information on these states. 
Again neglecting the slow q dependence of the envelope caused by polarization, $I_{ee}(q)$ in general has a period of $2Q$\,=\,$2\pi/d$, different from $I_{eo}(q)$. 
We distinguish three different cases: 
(i) $|c|\gg|a|^2$, (ii) $c$\,=\,0, and (iii) $0 < |c|\ll |a|^2$. 
(i) $|c|\gg |a|^2$ implies that the interference term between middle and outer sites dominates the modulation, 
which requires a dominant occupation of the middle site. 
Beyond occupation, the interference term depends on the phases of the states and 
on the matrix elements, which are also sensitive to polarization effects. 
Using the trigonometric identities $2\sin^2(x/2)=1-\cos(x)$ and $2\cos^2(x/2)=1+\cos(x)$, one can find the limiting behavior of $I_{ee}$ for large $|c|$. 
For $c\gg |a|^2$, the modulated part of the RIXS intensity roughly is given by $\cos^2(qd/2)$ and for $c\ll -|a|^2$ it becomes $\sin^2(qd/2)$. 
Qualitatively, this behavior persists even for small values of $|c|$ as long as $|c|/|a|^2\geq 2$, see Fig.\ \ref{fig:I_ee-function}(b). 
The $\sin^2(qd/2)$-type modulation agrees with the experimental observation at high energies, see Fig.\ \ref{fig:Ba3NbIr3O12_l_integrals}.
(ii) A special case is achieved for $c$\,=\,0, 
i.e., vanishing interference between middle and outer sites. 
This gives rise to $\cos^2(qd)$ behavior, i.e., the system again mimics a dimer with site distance $2d$ and period $1Q$, similar to the case of $I_{eo}$ but with different phase, see black curve in Fig.\ \ref{fig:I_ee-function}(b).  
(iii) Finally, $0 < |c|\ll |a|^2$ corresponds to a small but finite  interference term between the middle and outer sites, which yields secondary maxima. 
These still can be observed for, e.g., $c$\,=\,$|a|^2$ 
at $q$\,=\,$mQ$ with odd $m$, see dashed curves in Fig.\ \ref{fig:I_ee-function}(b).

These results promise that the measured $q$-dependent RIXS intensity  provides important information on the wave functions, as recently demonstrated for the tetrahedral clusters in  GaTa$_4$Se$_8$ \cite{Magnaterra2024}. 
However, the situation is more complex when states with different symmetry are close in energy. 
Consider, e.g., the sum of RIXS intensities with $I_{eo}(q)$\,$\sim$\,$\sin^2(qd)$ and $I_{ee}(q)$ as given by Eq.\ \eqref{eq:even}. 
The latter contains a $\cos^2(qd)$ term that competes with $I_{eo}(q)$
such that the $\sin^2(qd)$ character becomes visible only for sufficient  intensity. 
Also this observation will be relevant for the comparison of our experimental and theoretical results, see Sec.\ \ref{sec:comparison}.

\subsection{RIXS intensity for a single hole}

As a simple example that showcases the features found in the previous section, we address a face-sharing trimer with a single hole for strong spin-orbit coupling, zero trigonal field $\Delta$\,=\,0, and hopping ratio $f$\,=\,-1.
The energies are plotted in Fig.\ \ref{fig:single_hole_I_vs_dq_spin_orbit_summary}(a) and have been discussed above. Bonding and anti-bonding states are even under inversion (blue), the non-bonding states are odd (yellow).
Panels (b)-(f) depict the $q$-dependent RIXS intensity for all possible excitations.  
First, the two excitations (d) and (e) from the even ground state to the odd non-bonding states stand out by showing $\sin^2(qd)$ behavior with period $1Q$, as described by $I_{eo}(q)$, cf.\ Eq.\ \eqref{eq:odd}. 
Second, excitation (c) to the anti-bonding $j$\,=\,1/2 state agrees with dominant $\sin^2(qd/2)$ behavior as discussed above for $I_{ee}(q)$ with $c \ll 0$, i.e., strong occupation of the middle site. Finally, excitations to $j$\,=\,3/2 ((b),(f)) feature secondary maxima that are expected for $|c|<2|a|^2$. 
This occurs, e.g., for small occupation of the middle site $M_2$, but in this case it is due to destructive interference.
The finite background stems from summing the intensities for vertical and horizontal outgoing polarization that exhibit different $q$ dependence. 
Note that the relative strength of main and secondary maxima changes with $q$, reflecting the slow change of prefactors upon changing the scattering geometry and thus polarization.

\section{Comparison of theory and experiment}
\label{sec:comparison}

\subsection{Ground state of a single trimer}
\label{sec:groundstate}

The compound Ba$_4$NbIr$_3$O$_{12}$ previously has been discussed as a spin-liquid candidate \cite{Nguyen2019,Thakur2020,Bandyopadhyay2024gapless}. 
We therefore first discuss the ground state of a trimer with two holes. 
The Hamiltonian in Eq.\ \eqref{eq:hamiltonian} has seven parameters. 
We fix spin-orbit coupling $\lambda$\,=\,430\,meV and Hund's coupling $J_H$\,=\,330\,meV to realistic values for Ir oxides \cite{Gretarsson13,Revelli2019resonant,Revelli19BaCeIrO6}.
Large values are expected for the intra-orbital Coulomb interaction, 
$U$\,$\approx$\,1-2\,eV, and for the hopping $t$\,=\,$t_{a_{1g}}$\,=\,0.5\,-\,1\,eV \cite{Revelli2019resonant,Revelli2022} due to the short Ir-Ir distance. 
The ratio $f$\,=\,$t_{e_g^\pi}/t_{a_{1g}}$ is expected to be negative, 
roughly $f$\,=\,-1/2 \cite{Li2020soft}. 
Finally, the trigonal crystal-field splitting may reach values as large as a few hundred meV and may differ substantially between the inner and outer sites, $\Delta_m \neq \Delta_o$ \cite{Komleva2020three}. 
Given these physical constraints, the ground state of a single two-hole trimer is {\it always} a singlet and even under inversion, with the two holes filling a bonding orbital. This is intuitively evident in the non-interacting case depicted in Figs.\ \ref{fig:single-hole_hopping} and \ref{fig:single_hole_I_vs_dq_spin_orbit_summary}, but is stable also in the presence of strong correlations \cite{Komleva2020three}, as discussed below.
This already implies that a single trimer in zero magnetic field carries a vanishing quasimolecular moment, $J$\,=\,0. The experimental observation of a $q$-dependent RIXS intensity, modulated with periods $1Q$ and $2Q$, firmly establishes that the quasimolecular picture is applicable to Ba$_4$NbIr$_3$O$_{12}$. 
This predicts a nonmagnetic ground state for Ba$_4$NbIr$_3$O$_{12}$ as long as defects and interactions between trimers can be neglected.

In polycrystalline Ba$_4$NbIr$_3$O$_{12}$, the magnetic susceptibility has been reported to follow the Curie-Weiss law with a small magnetic moment of 0.8\,$\mu_B$ per 
trimer or 0.3\,$\mu_B$ per Ir site \cite{Nguyen2019,Bandyopadhyay2024gapless}. Similar values have been found in single crystals \cite{Thakur2020}. 
It has been argued that this magnetic moment, even though small, is too large to be caused by defects \cite{Nguyen2019}. 
From the refinement of powder x-ray diffraction data, 8-10\,\% of Nb-Ir site mixing have been claimed \cite{Bandyopadhyay2024gapless}.
A possible spin-liquid behavior of Ba$_4$NbIr$_3$O$_{12}$ has been discussed on the basis of the small magnetic moment, the absence of long-range magnetic order down to 0.05\,K, 
the specific heat, and $\mu$SR data \cite{Nguyen2019,Thakur2020,Bandyopadhyay2024gapless}. 
Before discussing how a small but finite magnetic susceptibility may emerge from $J$\,=\,0 trimers in an external magnetic field, cf.\ Sec.\ \ref{sec:singlet}, we first address the excitations of Ba$_4$NbIr$_3$O$_{12}$.

\begin{figure}[tb]
	\centering
	\includegraphics[width=\columnwidth]{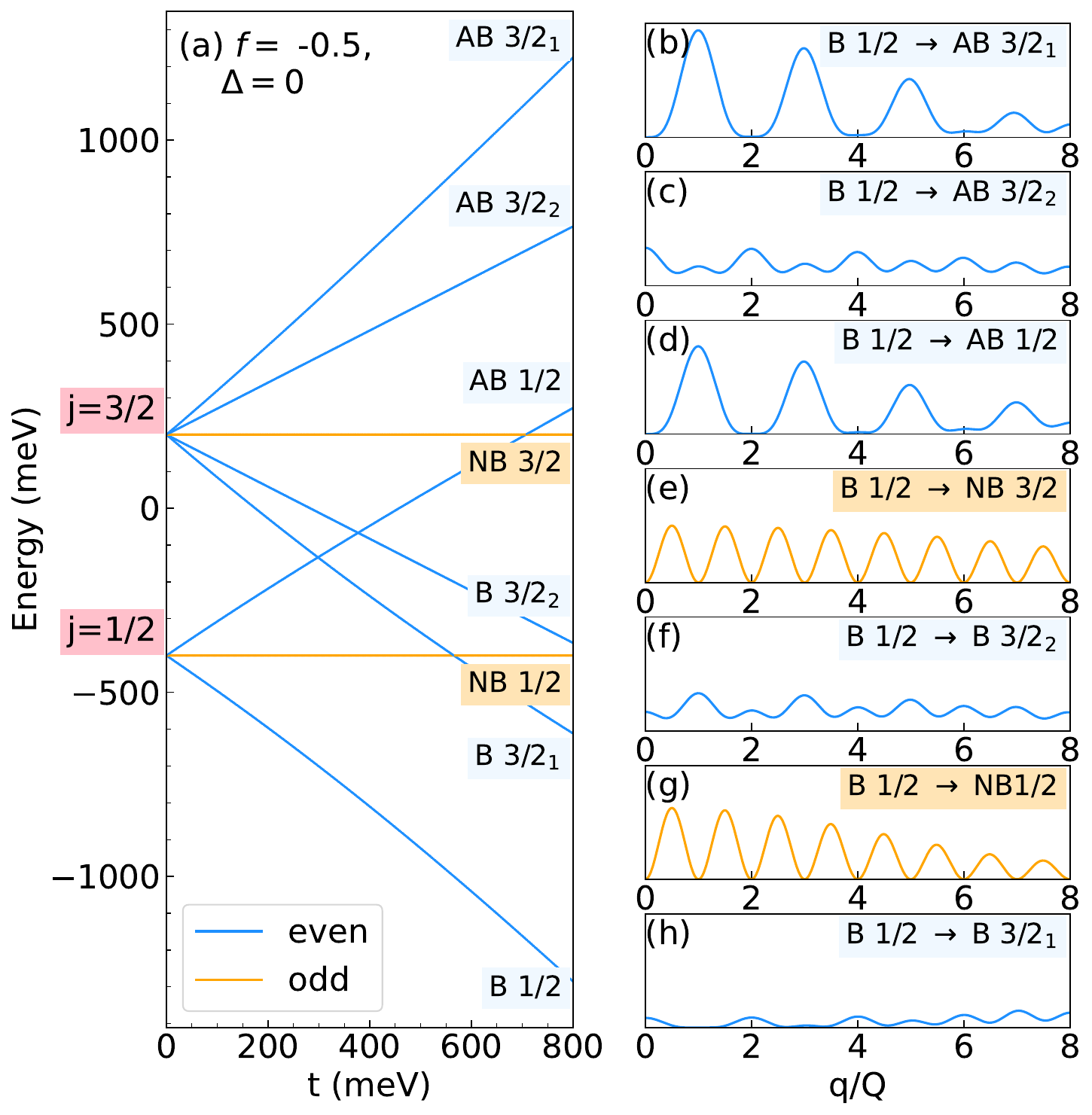}
	\caption{Quasimolecular orbital picture for strong spin-orbit coupling and vanishing trigonal field. Same as Fig.\ \ref{fig:single_hole_I_vs_dq_spin_orbit_summary} 
    but for realistic $f$\,=\,-0.5. The corresponding mixing of $j$\,=\,1/2 and 3/2 states lifts the degeneracy of bonding and anti-bonding $j$\,=\,3/2 quartets. 
    (a) Effect of hopping $t$ for $\lambda$\,=\,0.4\,eV.\@
    (b)-(h) RIXS intensity for a single hole per trimer. The panels show the $q$ dependence for excitations from the bonding $j$\,=\,1/2 ground state to each of the excited states. 
}
\label{fig:single_hole_I_vs_dq_spin_orbit_summary_f-0.5}
\end{figure}

\subsection{Excitations}
\label{sec:excitations}

For the discussion of the low-energy excitations, 
we start by summarizing the main features in the experimental data. 
A trimer exhibits a large number of eigenstates, in particular at higher energies, and the average over different contributions washes out the characteristic features. We thus will focus on the excitations below about 1.5\,eV, even though the overall behavior is similar up to 2\,eV.\@ 
(i) Up to 0.4\,eV, the RIXS intensity is negligible. 
(ii) At low energies, 0.5 -- 0.8\,eV, there is a clear $\sin^2(qd)$ behavior with period $1Q$, see Figs.\ \ref{fig:Ba3NbIr3O12_spectra_map} and \ref{fig:Ba3NbIr3O12_l_integrals}.
(iii) In the range 0.95 -- 1.55\,eV the dominant behavior is $\sin^2(qd/2)$ with period $2Q$.
On top, the integrated RIXS intensity, in particular from 0.95 to 1.25\,eV, exhibits shoulders at half-integer values of $Q$, indicating a small $\sin^2
(qd)$ contribution, see Fig.\ \ref{fig:Ba3NbIr3O12_l_integrals}.

\begin{figure*}[t]
	\centering
 \includegraphics[width=0.52\textwidth]{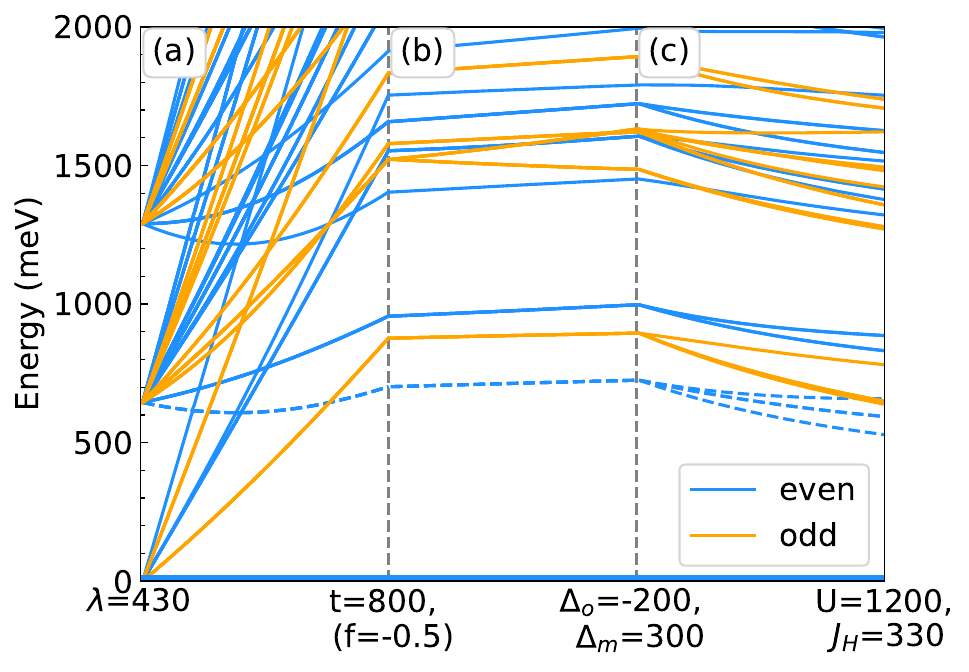}
 \includegraphics[width=0.45\textwidth]{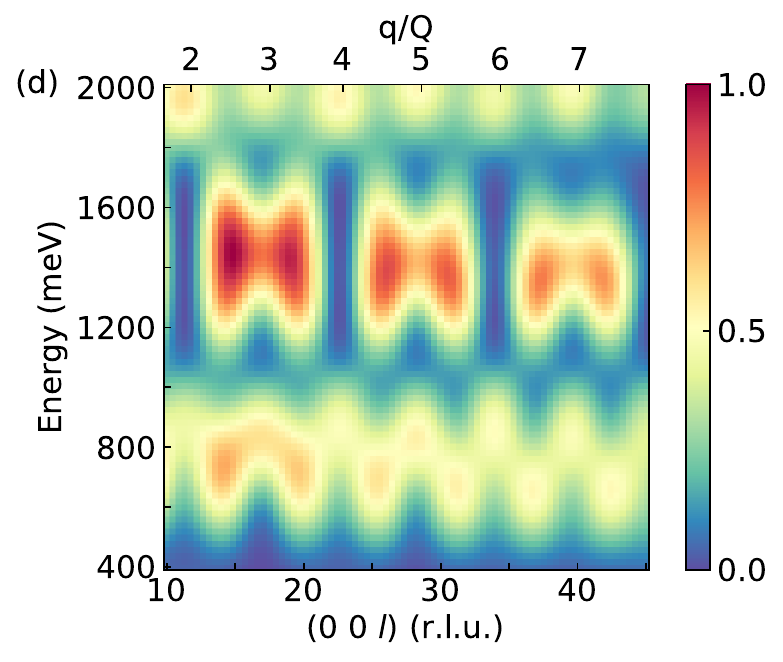}
	\caption{Left: Energies of the two-hole eigenstates of a face-sharing trimer, relative to the ground state at zero. Blue (yellow) denotes states that are even (odd) upon inversion.
    (a) Non-interacting case with spin-orbit coupling $\lambda$\,=\,0.43\,eV and vanishing trigonal crystal field, 
    $\Delta$\,=\,0.
    Hopping $t$\,=\,$t_{a_{1g}}$ increases linearly from zero to 0.8\,eV for fixed $f$\,=\,$t_{e_g^\pi}/t_{a_{1g}}$\,=\,-1/2. For $t$\,=\,0, both holes are in a $j$\,=\,1/2 state in the ground state, 
    and there are two possible excitation energies, 3/2\,$\lambda$ and 3\,$\lambda$, that correspond to one or two holes being excited to $j$\,=\,3/2, respectively. 
    In (b), $\Delta_o$ and $\Delta_m$ are turned on simultaneously. 
    In (c), Coulomb interactions are switched on, with $U$ and $J_H$ increasing up to 1.2\,eV and 0.33\,eV, respectively. 
    The full parameter set at the very right reads $\lambda$\,=\,0.43\,eV, $t$\,=\,0.8\,eV, $f$\,=\,-1/2, 
    $\Delta_o$\,=\,-0.2\,eV, $\Delta_m$\,=\,0.3\,eV, $U$\,=\,1.2\,eV, and $J_H$\,=\,0.33\,eV.\@
    The same parameters are used in panel (d), which shows the RIXS intensity on a color scale normalized to the maximum value.    
}
\label{fig:2holes_half_panel}
\end{figure*}

Observation (ii) of a $\sin^2(qd)$ modulation with period $1Q$ at low energy is well suited to narrow down the relevant parameter regime. 
It agrees with $I_{eo}(q)$, see Eq.\ \eqref{eq:odd}. 
Since the ground state is even, theory 
should yield a dominant inversion-odd state in the range 0.5 -- 0.8\,eV and no states with substantial RIXS intensity below. In the same spirit, this observation may serve to answer the question which, if any, non-interacting limit offers a simple, intuitive starting point to understand the full, interacting system. 
For instance, consider the non-interacting case studied in Fig.\ \ref{fig:single-hole_hopping} for vanishing spin-orbit coupling. 
A first excited state with odd symmetry (yellow) exists only for \textit{small} hopping $t$ and only if $\Delta_m\approx \Delta_o$, see Fig.\ \ref{fig:single-hole_hopping}(a). 
However, the lowest excited state is even under inversion (blue) for realistic, larger values of $t$, in particular for the realistic case of opposite signs of $\Delta_m$ and $\Delta_o$, see Fig.\ \ref{fig:single-hole_hopping}(b). 
Since this even-symmetry lowest excited state yields substantial RIXS intensity, the overall low-energy behavior of the non-interacting limit with $\lambda$\,=\,0 disagrees with the experimental findings.
In the following we will show that the correct low-energy behavior is obtained for realistic values of $\lambda$, highlighting the important role of spin-orbit coupling.

In addition to spin-orbit coupling, realistic hopping with 
$|t_{e_g^\pi}|$\,$<$\,$|t_{a_{1g}}|$, i.e., 0\,$>$\,$f$\,$>$\,-1 
is an essential ingredient to get a qualitatively correct $q$ dependence at low energies.
To see this, we first consider the simple, idealized case $f$\,=\,-1 which does not mix states with $j$\,=\,1/2 and $j$\,=\,3/2.  
We address the non-interacting case for $\lambda$\,=\,430\,meV and vanishing trigonal crystal field, see Fig.\ \ref{fig:single_hole_I_vs_dq_spin_orbit_summary}.
This scenario has the same problems as discussed above for the other limit, $\lambda$\,=\,0 (cf.\ Fig.\ \ref{fig:single-hole_hopping}): the lowest excited state for not too small hopping $t$ is even (blue curve) and thus exhibits the wrong symmetry. It gives a dominant $\cos^2(qd/2)$ behavior in the RIXS intensity, 
see Fig.\ \ref{fig:single_hole_I_vs_dq_spin_orbit_summary}(h), in contrast to the experimental low-energy result, 
$\sin^2(qd)$.
This first excited state is the quasimolecular equivalent of the spin-orbit exciton, i.e., the excitation of one of the holes from the $j$\,=\,1/2 bonding to the $j$\,=\,3/2 bonding multiplet at energy $E$\,=\,3/2$\lambda$. 
The discrepancy with experiment persists upon considering the trigonal crystal-field splitting and/or finite Coulomb interactions, even though the trigonal crystal field also mixes the $j$\,=\,1/2 and 3/2 multiplets. The idealized case $f$\,=\,-1 hence fails to describe the dominant $q$ dependence of the RIXS intensity.

However, the behavior is qualitatively different for realistic values of both $f$\,=\,-1/2 and $\lambda$\,=\,430\,meV, see Fig.\ \ref{fig:single_hole_I_vs_dq_spin_orbit_summary_f-0.5}. 
By mixing with the $j$\,=\,1/2 states, the $j$\,=\,3/2 multiplets split into two bonding doublets (blue), two anti-bonding doublets (blue), and a non-bonding quadruplet (yellow). 
For simplicity, we denote, e.g., the bonding doublets as bonding $j$\,=\,3/2$_1$ and 3/2$_2$. Strictly speaking, $j$ is not a good quantum number anymore, but most of the weight of a given wave function still lies in either $j$\,=\,3/2 or 1/2 states.
The essential point is that the lowest excitation to bonding $j$\,=\,3/2$_1$ carries nearly vanishing RIXS intensity, see Fig.\ \ref{fig:single_hole_I_vs_dq_spin_orbit_summary_f-0.5}(h).
At the same time, the higher bonding branch 
$j$\,=\,3/2$_2$ remains above the lowest inversion-odd non-bonding state, as long as the hopping $t$ does not become too large. 
Altogether, the $\sin^2(qd)$ behavior of the non-bonding branch (cf.\ Fig.\ \ref{fig:single_hole_I_vs_dq_spin_orbit_summary_f-0.5}(g)) dominates the RIXS intensity at low energy, in agreement with experiment.

In fact, this non-interacting scenario with strong spin-orbit coupling, large hopping $t$\,$\approx$\,0.8\,eV, and $f$\,=\,-1/2 already yields a qualitatively correct description of the three main experimental features. It has basically no RIXS intensity below 0.5\,eV, a dominant $\sin^2(qd)$ modulation with period $Q$ at low energies, and a dominant $\sin^2(qd/2)$ contribution with period $2Q$ at intermediate energies of about 1.5\,eV.\@ 
The latter originates from the excitation to anti-bonding $j$\,=\,1/2 states, see Figs.\ \ref{fig:single_hole_I_vs_dq_spin_orbit_summary_f-0.5}(a) and (d). 
Moreover, excitations to non-bonding $j$\,=\,3/2 states add enhanced intensity 
at half-integer $Q$ in the same intermediate energy range, see Fig.\ \ref{fig:single_hole_I_vs_dq_spin_orbit_summary_f-0.5}(e), and this can be identified with the 
shoulders at $(m+1/2)Q$ observed in our RIXS data, see Fig.\ \ref{fig:Ba3NbIr3O12_l_integrals}.

\begin{figure}[t]
	\centering
 \includegraphics[width=0.49\textwidth]{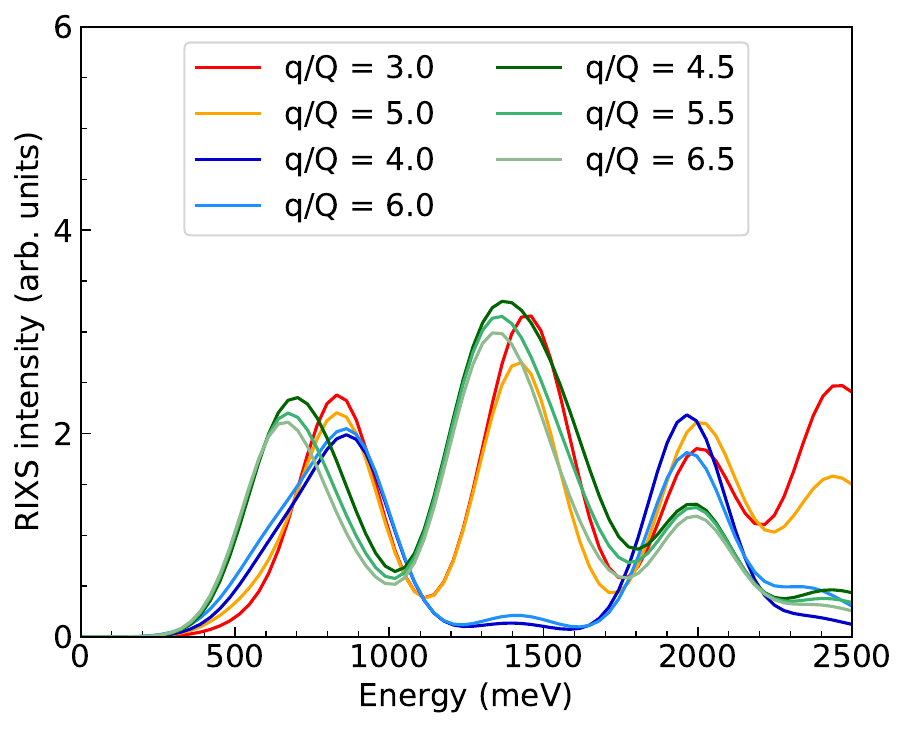}
	\caption{Calculated RIXS spectra of intra-$t_{2g}$ excitations of the two-hole trimer for different values of $q/Q$. 
    The parameters are the same as used in Fig.\ \ref{fig:2holes_half_panel}(d), i.e., 
    $\lambda$\,=\,0.43\,eV, $t$\,=\,0.8\,eV, $f$\,=\,-1/2, $\Delta_o$\,=\,-0.2\,eV, 
    $\Delta_m$\,=\,0.3\,eV, $U$\,=\,1.2\,eV, and $J_H$\,=\,0.33\,eV.
}
\label{fig:2holes_spectra}
\end{figure}

For a quantitative comparison between theory and experiment, we have to go beyond the scenario plotted in Fig.\ \ref{fig:single_hole_I_vs_dq_spin_orbit_summary_f-0.5} and include the trigonal crystal-field splitting as well as Coulomb interactions. Their effect on the two-hole energies is shown in Fig.\ \ref{fig:2holes_half_panel}. 
Panel (a) depicts the non-interacting case 
for $\Delta$\,=\,0 with $f$\,=\,\mbox{-1/2} and $\lambda$\,=\,430\,meV, as discussed above. It features an even ground state, and the lowest excited state for small $t$ shows  odd, non-bonding character (yellow). 
At $t$\,=\,800\,meV, the bonding $j$\,=\,3/2$_1$ state (blue dashed) is even lower in energy, but it carries very little RIXS intensity. 
At high energies, there is a multitude of states even in the non-interacting case. This makes it difficult to identify the dominant behavior, as mentioned above. Note that the states starting with an energy of $3\lambda$\,=\,1290\,meV at $t$\,=\,0 have both holes in the $j$\,=\,3/2 multiplet. For $t$\,=\,0, these states cannot be reached by the promotion of a single hole from the ground state in which both holes occupy $j$\,=\,1/2 states. This rule is not valid anymore for finite $t$ and interactions, but the RIXS intensity of these states remains small.

\begin{figure*}[t]
	\centering
 \includegraphics[width=0.98\textwidth]{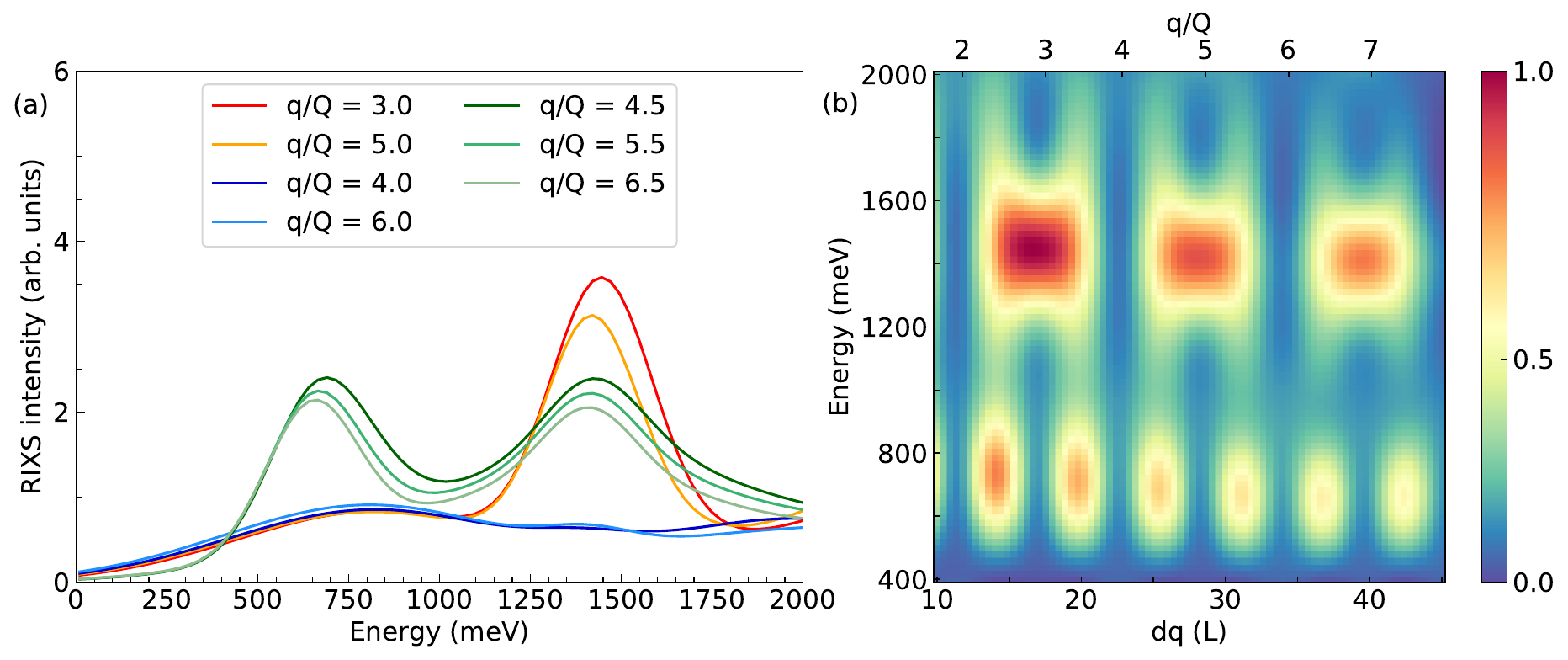}
 	\caption{(a) Calculated RIXS spectra and (b) $q$-dependent RIXS intensity for the same parameters as in Figs.\ \ref{fig:2holes_spectra} and \ref{fig:2holes_half_panel}(d). The single exception is the line width of excitations to $j$\,=\,3/2 type states, which here we choose to be 375\,meV, i.e., three times larger than for excitations to $j$\,=\,1/2 states. Overall, this yields good agreement with the experimental result, cf.\ Fig.\ \ref{fig:Ba3NbIr3O12_spectra_map}. 
    We propose that the enhanced line width reflects the interaction with continuum states above the Mott gap.
}
\label{fig:2holes_width}
\end{figure*}

In Fig.\ \ref{fig:2holes_half_panel}(b), the trigonal crystal-field splittings $\Delta_m$ and $\Delta_o$ simultaneously are tuned from zero on the left to $\Delta_m$\,=\,-200\,meV and $\Delta_o$\,=\,300\,meV on the right. 
For these realistic values, the trigonal crystal field has hardly any effect on the low-energy states. 
Similarly, panel (c) shows the increase of Coulomb interactions from zero to $U$\,=\,1200\,meV and $J_H$\,=\,330\,meV.\@ This causes a splitting of the multiplets and generically lowers the excitation energies. 
However, the qualitative behavior changes very little. 
We thus conclude that a substantial part of the excitation spectrum of the two-hole trimer can be motivated and discussed in a non-interacting picture as long as realistic values for $\lambda$, $t$, and $f$ are considered. Despite the large value of $U$, this may not come as a surprise. In cluster Mott insulators, the main role of a repulsive $U$ is to prevent charge fluctuations between different clusters, turning the entire system  insulating. In other words, $U$ suppresses  the small hopping between clusters. Within a given cluster, however, the large hopping strongly competes with Coulomb correlations and charge carriers are fully delocalized over the cluster. Furthermore, correlations are less relevant for the present case with less holes than sites. Together, this intuitively explains the success of the non-interacting scenario for the discussion of a single trimer.

The $q$-dependent RIXS intensity for the interacting system is shown as a heat map in Fig.\ \ref{fig:2holes_half_panel}(d), and the corresponding RIXS spectra for selected $q$ values are given in Fig.\ \ref{fig:2holes_spectra}. 
Both plots use the parameter set that corresponds to the right end of Fig.\ \ref{fig:2holes_half_panel}(c). 
The most important qualitative features of the experimental data, in particular (i)-(iii) mentioned at the beginning of Sec.\ \ref{sec:excitations}, are well reproduced. 
Note that the seven parameters in the Hamiltonian, Eq.\ \eqref{eq:hamiltonian}, span a large space, such that not all of them, in particular $\Delta_m$ and $\Delta_o$, can be determined accurately. 
However, these results of the interacting model support the general outcome of the non-interacting case: the physics of Ba$_4$NbIr$_3$O$_{12}$ is governed by strong spin-orbit coupling $\lambda$ of order 0.4\,eV, large hopping $t_{a_{1g}}$ of order 0.8\,eV, and 
$f$\,=\,$t_{e_g^\pi}/t_{a_{qg}}$\,$\approx$\,-1/2. 
Large hopping is required to achieve vanishing RIXS intensity at low energy, i.e., below 0.4\,eV, and the combination with strong $\lambda$ and $f$\,$\approx$\,-1/2 is necessary to reproduce the observed $1Q$-periodic intensity modulation of the low-energy RIXS peak. 
Accordingly, the ground state of a single trimer is well described by putting both holes in the quasimolecular bonding $j$\,=\,1/2 orbital, resulting in a total $J$\,=\,0 state 
with even symmetry. Concerning the spatial distribution of the two holes, 
we first consider $U$\,=\,$J_H$\,=\,$\Delta_m$\,=\,$\Delta_o$\,=\,0. In this simple case, we find one hole on the middle site while the outer sites carry half a hole each. For the realistic values of $U$, $J_H$, $\Delta_m$, and $\Delta_o$ discussed above, these numbers change by less than 10\,\%.

Our results challenge previously reported points of view on face-sharing iridate trimers. 
These have been discussed in terms of two limits, 
$\lambda$\,$\gg$\,$t$ or $t$\,$\gg$\,$\lambda$ \cite{Komleva2020three}. The first limit with extreme spin-orbit coupling considers local $j$\,=\,1/2 moments on individual Ir sites, where finite hopping yields exchange interactions between these local moments.
The compound Ba$_4$Ir$_3$O$_{10}$ with three holes per trimer or one hole per Ir site has been discussed in this limit \cite{Cao2020,Shen2022}.
The opposite limit of extreme hopping $t$\,$\gg$\,$\lambda$ considers quasimolecular orbitals that are built from $a_{1g}$ and $e_g^\pi$ orbitals \cite{Komleva2020three}, i.e., it also assumes $\Delta$\,$\gg$\,$\lambda$. 
Ye \textit{et al}.\ \cite{Ye2018covalency} claimed the covalency-driven collapse of strong spin-orbit coupling 
for Ba$_5$CuIr$_3$O$_{12}$ with three holes per trimer.
Based on Raman scattering and density functional theory, they discussed the local $j$\,=\,1/2 scenario against a quasimolecular-orbital scenario where spin-orbit coupling was applied only to the 
bonding $a_{1g}$ and $e_g^\pi$ states (in the hole picture).
Concerning Ba$_4$NbIr$_3$O$_{12}$, Bandyopadhyay \textit{et al}.\ \cite{Bandyopadhyay2024gapless} discussed several properties in a $j$\,=\,1/2 scenario but at the same time interpreted the small magnetic moment and the suppressed bandwidth as support for a quasimolecular picture. 
Analyzing the Ir $L_3$ and $L_2$ white-line intensities in x-ray absorption (XAS), they claim that spin-orbit coupling is suppressed to a moderate value. However, this analysis of the branching ratio employs a single-site picture, which is not appropriate for the trimer compound Ba$_4$NbIr$_3$O$_{12}$, in particular not for a quantitative analysis. 
Our results highlight the quasimolecular nature of the trimer states but at the same time emphasize the important role of spin-orbit coupling, despite the large hopping.
For a quantitative description, one has to treat spin-orbit coupling, hopping, Coulomb interactions, and the trigonal crystal field on equal footing. However, we have shown that a qualitative understanding can be achieved by considering quasimolecular orbitals built from spin-orbit-entangled $j$ states but not from $a_{1g}$ or $e_g^\pi$ states 
for $\lambda$\,=\,0.

Finally, we comment on the shortcomings of our theoretical approach. 
Firstly, the RIXS data at 1-1.5\,eV show a dominant $2Q$-periodic $\sin^2(qd/2)$ modulation and additionally weak shoulders at half-integer $Q$, i.e., a weak $\sin^2(qd)$ contribution, see Fig.\ \ref{fig:Ba3NbIr3O12_l_integrals}. 
The former is well described by the excitation to the quasimolecular anti-bonding $j$\,=\,1/2 state, see Fig.\ \ref{fig:single_hole_I_vs_dq_spin_orbit_summary_f-0.5}(d). 
However, theory overestimates the intensity of the $1Q$-periodic features, i.e., the $\sin^2(qd)$ contribution of excitations to non-bonding $j$\,=\,3/2 states, cf.\ Figs.\ \ref{fig:single_hole_I_vs_dq_spin_orbit_summary_f-0.5}(e) and \ref{fig:2holes_half_panel}. 
Secondly, just below 1\,eV theory shows a predominant $\cos^2(qd)$\,=\,$\cos^2(\pi q/Q)$ behavior with maximum intensity for integer $Q$, out-of-phase with the $\sin^2(qd)$ modulation around 0.7\,eV.\@ 
In contrast, our experimental data show a clear minimum for $4Q$ and $6Q$, see red and black curves in Fig.\ \ref{fig:Ba3NbIr3O12_l_integrals}.
Also this shortcoming can be viewed as an overestimation of excitations to $j$\,=\,3/2 states, in this case to bonding orbitals, see Fig.\ \ref{fig:single_hole_I_vs_dq_spin_orbit_summary_f-0.5}(f) and (h).

To find a possible reason for the overestimated intensity of excitations to $j$\,=\,3/2 states, we discuss the assumptions employed in our theoretical approach.
First, we neglect the small admixture of $e_g^\sigma$ orbitals. 
With the cubic crystal-field splitting 10\,$Dq$\,$\approx$\,3.5\,eV,
we expect this to have a minor effect on the intra-$t_{2g}$ excitations below 2\,eV, 
see Fig.\ \ref{fig:eg}. 
Second, we neglect the possible dynamics in the intermediate state of the RIXS process. This is very well justified for a single Ir site, where the Ir $2p^5 t_{2g}^6$ intermediate state has a full $t_{2g}$ shell. 
For a trimer, there are different intermediate states with one hole in the quasimolecular $t_{2g}$ orbitals (and the second one in a $2p$ core state).
However, the large width of a few eV of the RIXS resonance curve and its featureless line shape 
suggest that all intermediate states contribute equally to the RIXS amplitude. 
Using this assumption, the dynamics in the intermediate state can be neglected.
Third, we consider a single trimer and ignore interaction effects with the electronic continuum above the Mott gap, i.e., \textit{inter}-trimer excitations.
The actual size of the gap is not known well thus far. The activation energy $\Delta_{\rm act}$ determined from the electrical resistivity may serve as a lower limit. 
Values of $\Delta_{\rm act}$\,=\,0.05\,eV \cite{Nguyen2019,Thakur2020} and 0.22-0.25\,eV \cite{Bandyopadhyay2024gapless} have been reported. 
Typically, the size of the Mott gap cannot be determined from RIXS at the transition-metal $L$ edge \cite{Benckiser2013}. With a $2p$ core hole in the intermediate state, the RIXS process in good approximation can be viewed as a coherent superposition of single-site scattering events, see Eq.\ \eqref{eq:RIXS_amplitude}. 
It is nevertheless possible that the RIXS spectra exhibit a weak, broad, and featureless contribution of inter-trimer excitations across the Mott gap. In particular, the small RIXS intensity of the latter may be enhanced in the case of hybridization between intra-trimer and inter-trimer excitations. 
Our single-trimer theory predicts a nearly vanishing RIXS intensity just below 1.5\,eV for $q$\,=\,$4Q$ or $6Q$, see Fig.\ \ref{fig:2holes_spectra}. 
In contrast, the RIXS data between 1 and 1.5\,eV show a weak  background that indeed may indicate a finite contribution of inter-trimer excitations (see blue curves in Fig.\ \ref{fig:Ba3NbIr3O12_spectra_map}). 
A way to explain both the overestimated intensity of excitations to $j$\,=\,3/2 type of states and a finite continuum contribution is to assume a certain hybridization between such excitations to $j$\,=\,3/2 states and electronic continuum states.  
To mimic this, we empirically assume a line width of 375\,meV for excitations to $j$\,=\,3/2 states, which is three times larger than for $j$\,=\,1/2. In this case, the calculated result in fact shows good agreement with the experimental data, see Figs.\ \ref{fig:2holes_width} and \ref{fig:Ba3NbIr3O12_spectra_map}.

\subsection{Magnetism vs.\ $J$\,=\,0}
\label{sec:singlet}

From our analysis of a single trimer, the ground state is very stable and given by a singlet. This raises the question of the origin of the Curie-Weiss contribution observed in the magnetic susceptibility $\chi(T)$, corresponding to a small magnetic moment of about 0.3\,$\mu_B$ per Ir site \cite{Nguyen2019,Thakur2020,Bandyopadhyay2024gapless}. 
Excitonic magnetism emerging from a nonmagnetic ground state has been discussed for, e.g., $d^4$ $J$\,=\,0 systems \cite{Khaliullin2013,Khomskii_transition_2014}. 
It arises in the case of condensation of a magnetic excited state that exhibits large dispersion \cite{Khaliullin2013,Kaushal2021}. In other words, the system flips to an excited magnetic state on every site since the energy cost is overruled by the gain of exchange energy. 
Such condensation yields an entirely different ground state, which we rule out for Ba$_4$NbIr$_3$O$_{12}$ based on the good agreement between our experimental and theoretical results. 
Moreover, excitonic magnetism requires strong inter-trimer exchange interactions and a corresponding pronounced dispersion that, however, is absent in our RIXS data.

Furthermore, spin-orbit coupling causes Van Vleck paramagnetism in an external magnetic field since the magnetization operator mixes the zero-field eigenstates. This may give rise to the admixture of a magnetic state to the ground state, as discussed, e.g., for the $5d^4$ $J$\,=\,0 ground state of K$_2$OsCl$_6$ \cite{Warzanowski2023}.
We consider local excited states of individual, noninteracting trimers as well as excitations across the Mott gap that may admix states with different hole count via \textit{inter}-trimer hopping. 
The lowest excitation energy of a single trimer amounts to about 0.5\,eV, and a lower limit for intersite excitations is given by the activation energy 0.25\,eV \cite{Bandyopadhyay2024gapless}, as discussed above. 
Given these large excitation energies, the Van Vleck contribution to $\chi(T)$ is certainly small and nearly independent of temperature below 300\,K.\@ 
Altogether, we expect a small Van Vleck contribution in $\chi(T)$ of Ba$_4$NbIr$_3$O$_{12}$ but rule out that the trimers with two holes cause a finite Curie-Weiss contribution in a small magnetic field.

\section{Conclusion}

In conclusion, we find that the RIXS spectra and in particular the momentum dependence of the RIXS intensity $I(\mathbf{q})$ of Ba$_4$NbIr$_3$O$_{12}$ yield clear fingerprints of the quasimolecular electronic structure of this trimer compound. 
In insulating Ba$_4$NbIr$_3$O$_{12}$, the two holes per trimer are delocalized in quasimolecular trimer orbitals.
One characteristic feature of a trimer is the existence of two different periods in $I(\mathbf{q})$ that reflect the intra-trimer Ir-Ir distances $d$ and $2d$. Beyond the two periods, the presence of inversion symmetry for Ir$_3$O$_{12}$ trimers in face-sharing geometry yields characteristic interference patterns in $I(\mathbf{q})$ that facilitate the interpretation of the RIXS spectra. This allows us to determine a realistic range of parameters. Remarkably, a non-interacting picture with strong spin-orbit coupling $\lambda$\,$\approx$\,0.4\,eV and large hopping $t_{a_{1g}}$\,$\approx$\,0.8\,eV with $t_{e_g^\pi}/t_{a_{1g}}$\,$\approx$\,-1/2 already describes the main qualitative features of the RIXS data. 
In this regime, a trigonal crystal field of up to a few hundred meV has only a small effect on the RIXS response and does not seem essential to understand the qualitative physics. 
Concerning Coulomb interactions, we employ $U$\,=\,1.2\,eV and $J_H$\,=\,0.33\,eV, but also these are not essential for a qualitative description of the properties of a single trimer.  
On a trimer, hopping wins against correlations, in particular for a hole count smaller than the number of Ir sites.

In the literature, the physics of $5d$ iridate trimers has been discussed in two limits, either extreme spin-orbit coupling with local $j$\,=\,1/2 moments or extreme hopping, i.e., a quasimolecular-orbital picture based on $a_{1g}$ and $e_g^\pi$ orbitals  \cite{Komleva2020three,Cao2020,Shen2022,Ye2018covalency,Bandyopadhyay2024gapless}. 
In the quasimolecular scenario, spin-orbit coupling has either been neglected or applied only to the bonding states \cite{Ye2018covalency}. We find that a quantitative description requires to consider spin-orbit coupling and hopping on the same footing and we show that the trimer physics of Ba$_4$NbIr$_3$O$_{12}$ can be understood very well in terms of quasimolecular states that are formed from spin-orbit-entangled $j$ moments. This agrees with previous RIXS results on iridate dimers \cite{Revelli2019resonant,Revelli2022}. 
Remarkably, a different scenario has been found in RIXS on Ta tetrahedra in GaTa$_4$Se$_8$, where an intuitive picture is obtained by first considering hopping and then applying spin-orbit coupling only to the quasimolecular $t_2$ orbitals \cite{Magnaterra2024}. The choice of an appropriate intuitive model hence depends on the cluster shape and the relative size of electronic parameters. 
The essential role of spin-orbit coupling in these cluster Mott insulators promises a non-trivial character of the magnetic moments in, e.g., trimers with an odd number of holes. In general, theoretical investigations of exchange interactions between neighboring clusters are highly desirable.

For realistic parameters, we show that the ground state of a trimer with two holes is a nonmagnetic singlet that is even under inversion. 
Both holes occupy the bonding $j$\,=\,1/2 orbital, which yields a total $J$\,=\,0.
A nonmagnetic ground state has also been obtained in density-functional calculations for Ba$_4$NbIr$_3$O$_{12}$ \cite{Komleva2020three}. Actually, the insulating character arises as soon as realistic spin-orbit coupling is included, i.e., even without correlations, which has been rationalized by the two holes filling the lowest cluster orbital \cite{Komleva2020three}.
This raises the question on the classification of Ba$_4$NbIr$_3$O$_{12}$ as either a cluster Mott insulator or a cluster-type band insulator. 
Our RIXS data show that the excited states can be assigned to a given trimer, supporting a significant role of inter-trimer Coulomb interactions and a cluster Mott picture.

\begin{acknowledgments}
Our research project on cluster Mott insulators to a large extent has been triggered by fruitful discussions with Daniel I. Khomskii, to whom we pay our heartfelt tribute. 
We acknowledge the European Synchrotron Radiation Facility (ESRF) for 
providing beam time at beamline ID20 under proposal number IH-HC-3879 and technical support.
Furthermore, we acknowledge funding from the Deutsche Forschungsgemeinschaft 
(DFG, German Research Foundation) through Project No.\ 277146847 -- CRC 1238 
(projects A02, B03) 
as well as from the European Union - Next Generation EU - “PNRR-M4C2, investimento 1.1-Fondo PRIN 2022” - “Superlattices of relativistic oxides” (ID No.\ 2022L28H97, CUP D53D23002260006).
A.S.\@ and M.H.\@ thank J. Attig for numerical support and acknowledge partial funding by the Knut and Alice Wallenberg Foundation as part of the Wallenberg Academy Fellows project. 
\end{acknowledgments}

\appendix

\section{Sample characterization}
\label{App:sample}

For crystals grown in Cologne, the chemical composition and homogeneity were determined with a JEOL JXA-8900RL Electronbeam Microprobe. 
On a polished planar surface, the Ba and Nb concentration were measured using a PET crystal spectrometer with baryte and elemental Nb as standards. The Ir concentration was measured using a LiF crystal spectrometer with an IrO$_2$ crystal as standard.

\section{Details on the RIXS intensity}
\label{sec:Intensity_inversion}

Here, we review how to derive Eq.\ \eqref{eq:odd}.
A generic state on the trimer can be written as 
	\begin{align}
		|\psi\rangle &=\sum_{ijk}c_{ijk}|i;j;k\rangle
	\end{align}
where $i,j,k$ denote the basis states on the sites $M_1$, $M_2$, and $M_3$, 
see Fig.\ \ref{fig:struc}.
In general, $i,j,k$ denote many-body states and need not have fixed particle number individually. Only the total particle number is fixed.
How the single-particle orbitals transform under inversion depends on the material at hand. 
For the trimers in Ba$_4$NbIr$_3$O$_{12}$, the single-particle basis states --- and consequently the many-particle ones as well --- transform in a very simple way under inversion,
\begin{align}
		\hat I |i;j;k\rangle &= |k;j;i\rangle . 
	\end{align}
	Thus, inversion constrains the coefficients $c_{ijk}$ to 
	\begin{align}
		c_{ijk}&=\pm c_{kji}, 
		\label{eq:Inv}
\end{align}
where the $+$ ($-$) sign denotes even (odd) states under inversion. 
For instance, bonding and anti-bonding states are both inversion-symmetric (even), while non-bonding states are anti-symmetric (odd). 

For a single trimer, the RIXS amplitude for generic initial and final states $|\psi_i\rangle$ and $|\psi_f\rangle$ reads (cf.\ Eq.\ \eqref{eq:RIXS_amplitude})
\begin{align}\label{eq:app_Afirst}
	A(\mathbf{q})&\sim \sum_{\mathbf R} e^{i \mathbf{q}\mathbf{R}} \langle \psi_f | \left[D^\dagger(\epsilon_{\rm out}^*)D(\epsilon_{\rm in})\right]_\mathbf{R} |\psi_i\rangle \nonumber\\
	&= e^{iqd} \sum_{i,j,k}\sum_{i',j',k'} \tilde c^*_{i'j'k'}c_{ijk} \delta_{j,j'}\delta_{k,k'} \langle i' |D^\dagger D| i\rangle_1\nonumber\\
	&+ e^{-iqd} \sum_{i,j,k}\sum_{i',j',k'} \tilde c^*_{i'j'k'}c_{ijk} \delta_{j,j'}\delta_{i,i'} \langle k' |D^\dagger D| k\rangle_3 \nonumber\\
	&+  \sum_{i,j,k}\sum_{i',j',k'} \tilde c^*_{i'j'k'}c_{ijk} \delta_{i,i'}\delta_{k,k'} \langle j' |D^\dagger D| j\rangle_2  
\end{align}	
where the subscript of $|\ldots \rangle_m$ denotes the site. 
We can combine the terms in the second and third lines, 
using that the RIXS amplitudes for the outer sites are identical due to inversion symmetry, i.e.\ 
$ \langle j |D^\dagger D| i\rangle_1 =   \langle j |D^\dagger D| i\rangle_ 3$. 
To obtain a compact expression, we rename the dummy indices labeling the states on each site (e.g. in line three, we change indices as $k\rightarrow i$, $k'\rightarrow j$, $j=j' \rightarrow \alpha$ and $i=i' \rightarrow \beta$).   
This yields 
\begin{align}\label{eq:app_A}
 A(\mathbf{q})&\sim
 \sum_{i,j,\alpha,\beta} \! \Big[(\tilde c^*_{j\alpha\beta}c_{i\alpha\beta} e^{i qd} +\tilde c^*_{\beta\alpha j}c_{\beta\alpha i} e^{-iqd}) 
   \cdot 
   \langle j |D^\dagger D| i\rangle_1  \nonumber\\
   &+ \tilde c^*_{\alpha j \beta} c_{\alpha i \beta } 
   \langle j |D^\dagger D| i\rangle_2    \Big]
\end{align}	
Note, that the RIXS amplitude on the middle site $M_2$ may differ from the ones on the outer sites and, thus, cannot be combined with the other two terms. 
This is, e.g., the case for a face-sharing trimer, where the middle octahedron is rotated by 180$^\circ$ around the global $z$ axis compared to the outer ones.

If the initial and final states have different symmetry, one may use $\tilde c^*_{\beta \alpha j}c_{\beta \alpha i} = - c^*_{j\alpha \beta }c_{i \alpha \beta}$ to simplify the expression further, 
\begin{align}\label{eq:app_A(q)even_to_odd}
 A_{eo}(\mathbf{q}) 
 & \sim \sum_{i,j,\alpha,\beta} \tilde c^*_{j\alpha\beta}c_{i\alpha\beta }\underbrace{(e^{i q d} - e^{-iq d})}_{2i\sin(q d)} \cdot   
 \langle j |D^\dagger D| i\rangle_1 ,
\end{align}
where the contribution from the middle site vanishes under summation over $\alpha$ and $\beta$. The full intensity for an excitation that flips the symmetry is then given by 
\begin{align}\label{eq:app_odd}
 I_{eo}(q)
 &\sim 4\sin^2(qd) \cdot \Big| \sum_{i,j,\alpha\beta} \tilde c^*_{j\alpha\beta }c_{i\alpha\beta}\,  
 \langle j |D^\dagger D| i\rangle_1    \Big|^2, 
\end{align}
as claimed in the main text, see Eq.\ \eqref{eq:odd}.

For excitations between states with the same symmetry, we also can combine the terms from the outer sites (now with a plus sign), but the contribution from the middle site will not vanish (the only difference is that $c_{\alpha j\alpha}$ is required to vanish for odd states),  
\begin{align}\label{eq:app_even}
 A_{ee}(\mathbf{q})
 &=\sum_{i,j,\alpha,\beta} \tilde c^*_{j\alpha\beta}c_{i\alpha\beta }\, 2 \cos(qd) \cdot   
 \langle j |D^\dagger D| i\rangle_1   \nonumber\\
 &+ \tilde c^*_{\alpha j \beta} c_{\alpha i \beta } \,  
 \langle j |D^\dagger D| i\rangle_2    \nonumber\\
	&=  \cos(qd)\underbrace{\sum_{i,j,\alpha,\beta} 2 \tilde c^*_{j\alpha\beta}c_{i\alpha\beta }  
  \langle j |D^\dagger D| i\rangle_1     }_a \nonumber\\
 &+\underbrace{\sum_{i,j,\alpha,\beta}\tilde c^*_{\alpha j \beta} c_{\alpha i \beta }    
 \langle j |D^\dagger D| i\rangle_2   }_{b}. 
 \end{align}
Note that $a$ and $b$ may have a different dependence on $q$, even in the case where the dipole matrix elements are the same for all three sites, since the summation over the eigenstates may differ for outer and inner sites. 
From the amplitude in Eq.\ \eqref{eq:app_even}, it immediately follows that the RIXS intensities for excitations between states with the same symmetry is given as claimed in Eq.\ \eqref{eq:even}, 
\begin{align}
I_{ee}(q) 
&\sim |a|^2 \cos^2(qd) + (a^*b+b^* a) \cos(qd) + |b|^2\, .
\end{align}

%

\end{document}